\documentclass{article}

\usepackage{PRIMEarxiv}
\usepackage{natbib}
\usepackage{enumitem}
\usepackage{float}

\usepackage[utf8]{inputenc} % allow utf-8 input
\usepackage{amsmath}
\usepackage[T1]{fontenc}    % use 8-bit T1 fonts
\usepackage{hyperref}       % hyperlinks
\usepackage{url}            % simple URL typesetting
\usepackage{booktabs}       % professional-quality tables
\usepackage{amsfonts}       % blackboard math symbols
\usepackage{nicefrac}       % compact symbols for 1/2, etc.
\usepackage{microtype}      % microtypography
\usepackage{lipsum}
\usepackage{fancyhdr}       % header
\usepackage{graphicx}       % graphics
\graphicspath{{media/}}     % organize your images and other figures under media/

\title{
CAMulator: Fast Emulation of the Community Atmosphere Model
    \thanks{\textit{Citation}: Chapman W. et al. fully resolved, fast emulation of the Community Atmosphere Model}
    \author{
      \textbf{William E. Chapman}$^{\ddagger}$,
      \textbf{John S. Schreck}$^{\dagger}$,
      \textbf{Yingkai Sha}$^{\dagger}$,
      \textbf{David John Gagne II}$^{\dagger}$,
      \textbf{Dhamma Kimpara}$^{\dagger}$,
      \\
      \textbf{Laure Zanna}$^{\zeta}$,
      \textbf{Kirsten J. Mayer}$^{\ddagger}$,
      \textbf{Judith Berner}$^{\ddagger}$\\[1em]
      Climate and Global Dynamics (CGD) Laboratory$^{\ddagger}$\\
      Computational and Information Systems (CISL) Laboratory$^{\dagger}$ \\
      Courant Institute of Mathematical Sciences, New York University$^{\zeta}$\\[1em]
      NSF National Center for Atmospheric Research \\
      Boulder, Colorado, USA\\[1em]
      \texttt{\{wchapman, schreck, ksha\}@ucar.edu}
    }
}

\begin{document}
\maketitle

\begin{abstract}
We introduce CAMulator version 1, an auto-regressive machine-learned (ML) emulator of the Community Atmosphere Model version 6 (CAM6) that simulates the next atmospheric state given prescribed sea surface temperatures and incoming solar radiation. CAMulator explicitly conserves global dry air mass, moisture, and total atmospheric energy while remaining numerically stable over indefinite climate integrations. It successfully reproduces the annual CAM6 climatology and key modes of climate variability, including the El Ni\~{n}o–Southern Oscillation, the North Atlantic Oscillation, and the Pacific-North American pattern, with slightly muted variability. When forced with sea surface temperature (SST) beyond the training distribution, CAMulator exhibits a systematic cold bias in high-latitude regions, particularly in boreal winter, likely due to the absence of interactive land and sea ice. Nonetheless, CAMulator achieves these results with a 350 times speedup over CAM6, making it an efficient alternative for generating large ensembles. CAMulator represents a step toward fast, physically realistic ML-driven climate modeling.
\end{abstract}

\section{Introduction}

Earth system models are essential for understanding Earth system dynamics, characterizing extreme events, and projecting future climate scenarios. However, their high computational cost limits their practical applications. One major computational bottleneck of these models is the atmosphere component \citep{Worley2011}. Recent advances in machine learning (ML) emulators offer a promising path forward, with the potential to accelerate simulations and improve physical representation. Demonstrating the fidelity of ML-driven emulation is crucial; success in Earth system model component emulation could enhance and enable key modeling activities such as rapid hypothesis testing, large-ensemble generation \citep{kay2015community, maher2021large}, and comprehensive uncertainty quantification \citep[e.g.,][]{deser2017northern, touma2022climate}, ultimately broadening the scope and efficiency of climate research.

Progress in ML-based climate emulation has been rapid, particularly in emulating component models \citep{dheeshjith2024samudra, watt2023ace, watt2024ace2} and, more recently, in coupled systems \citep{cresswell2024deep}. The community is beginning to demonstrate that ML-based Earth system component models have the potential to emulate key features that are important for climate research, such as emergent variability and response to external forcing. However, the extent to which this has been achieved remains an active area of investigation \citep{watt2024ace2}. However, differences in the state variables, reduced dimensionality, and lower physical resolution—both horizontally and vertically—pose challenges for combining these ML-based atmospheric emulators with other Earth system components, including ocean, land, cryosphere and chemistry.It has also been proposed to interactively combine such emulators with dynamical models using a supermodeling framework \citep[e.g.,][]{schevenhoven2023supermodeling, chapman2024implementation}. These differences hinder the seamless coupling with traditional physics-based models and limit the broader adoption of  atmospheric ML-emulators in Earth system simulations.

Here, we introduce CAMulator Version 1, an ML-based emulator that mimics the Community Atmosphere Model version 6 (CAM6) \citep{danabasoglu2020community} while preserving both vertical and horizontal spatial resolution. CAMulator runs approximately 350 times faster than CAM6 (see Section~\ref{scomp} for details) while enforcing key conservation properties, including global dry mass, total water, and energy. 

For model training and testing, we leverage the National Science Foundation's (NSF) National Center for Atmospheric Research (NCAR)-Community Research Earth Digital Intelligence Twin (CREDIT) platform \citep{sha2025improving, schreck2024community}. CREDIT is a machine learning platform for scientific research that provides an efficient framework for the rapid development of auto-regressive models, making it well-suited for Earth system modeling tasks. In our study, CREDIT serves as the backbone for an efficient data pipeline, scalable high-performance computing, model training, and inference. We extend its capabilities from numerical weather prediction emulation to fast climate emulation and demonstrate the necessity of incorporating physical constraints on mass, water, and energy to enhance model fidelity and improve emulation accuracy.

In this work, we focus on emulating atmospheric physics during the historical period (1979–2014), a common validation period for CAM6. We conduct this experiment as an AMIP simulation, where CAMulator is forced with observed sea surface temperatures (SSTs) and incoming solar radiation. We show that CAMulator successfully captures the historical climate trend over this period. To test model stability in an indefinite forcing mode, we run long-term integrations using climatological SSTs as the prescribed boundary condition, ensuring consistency with the statistical properties of the CAM6 climate system.

We find that CAMulator successfully reproduces key climate statistics, including variability associated with the El Nino Southern Oscillation and has similar root mean square error compared to CAM6. Although geopotential height is not directly predicted, the emulator is still able to accurately capture major modes of variability, such as the North Atlantic Oscillation and the Pacific North American pattern, from predicted variables. This suggests that the model preserves physical consistency across key atmospheric fields, even in extended simulations.

An important application of climate models is to predict the response to an external forcing, for example increased ocean sea surface temperatures (SSTs). Beyond historical validation, we test CAMulator in out-of-sample scenarios (i.e. forcing scenarios the emulator was not trained on) with ocean surface warming of +2K and +4K,. The model demonstrates a promising ability to adjust dynamically to these warmer conditions, though its response appears weaker under +4K warming. In this manuscript, we focus on evaluating CAMulator’s performance and exploring its potential applications and extensions.

In this manuscript, we focus on evaluating CAMulator’s performance and exploring its potential applications and extensions and is organized as follows: Section 2 describes the data sources used for model training and evaluation. Section 3 outlines the methodology, including model construction and training. Section 4 presents the results, highlighting CAMulator’s performance in reproducing key climate statistics and its response to out-of-sample warming scenarios. Finally, we conclude with a discussion of CAMulator’s strengths, limitations, and potential extensions for future work.

\begin{table}[htbp]
    \centering
    \caption{Description of input and predicted variables for CAMulator. Variables are categorized as input-only (static and dynamic forcing), input/output (prognostic), and output-only (diagnostic). \textbf{$^*$}Qtot is a sum of all column moisture—both vapor and condensed: Qtot = Specific Humidity + Grid Box Snow Amount + Grid Box Rain Amount. TOA = Top of Atmosphere.}
    \label{tab:ml_variables}
    \begin{tabular}{llclc}
        \toprule
        \textbf{Variable} & \textbf{Description} & \textbf{Units} & \textbf{Levels} & \textbf{Category} \\
        \midrule
        \multicolumn{5}{c}{\textbf{Input Variables (Forcing)}} \\
        \midrule
        \multicolumn{5}{c}{\textit{Static Forcing (Time-Invariant Input)}} \\
        \midrule
        Surface Geop. & Surface Height & m²/s² & Single Level & Input \\
        Land-Sea Mask & Land Mask & Unitless & Single Level & Input \\
        \midrule
        \multicolumn{5}{c}{\textit{Dynamic Forcing (Time-Varying Input)}} \\
        \midrule
        SOLIN   & Incoming Solar Radiation & J/m² & Single Level & Input \\
        SST     & Sea Surface Temperature & K & Single Level & Input \\
        \midrule
        \multicolumn{5}{c}{\textbf{Prognostic Variables (Input and Output)}} \\
        \midrule
        U       & Zonal Wind & m/s & 32 Levels & Input/Output \\
        V       & Meridional Wind & m/s & 32 Levels & Input/Output \\
        T       & Temperature & K & 32 Levels & Input/Output \\
        \textbf{$^*$}Qtot  & Specific Total Water & kg/kg & 32 Levels & Input/Output \\
        PS      & Surface Pressure & Pa & Single Level & Input/Output \\
        TREFHT  & Near-Surface Air Temperature & K & Single Level & Input/Output \\
        \midrule
        \multicolumn{5}{c}{\textbf{Diagnostic Variables (Output Only)}} \\
        \midrule
        PRECT   & Accumulated Precipitation & m & Single Level & Output \\
        CLDTOT  & Total Cloud Cover & Fraction & Single Level & Output \\
        CLDHGH  & High Cloud Cover & Fraction & Single Level & Output \\
        CLDLOW  & Low Cloud Cover & Fraction & Single Level & Output \\
        CLDMED  & Medium Cloud Cover & Fraction & Single Level & Output \\
        TAUX    & Zonal Wind Stress & N/m² & Single Level & Output \\
        TAUY    & Meridional Wind Stress & N/m² & Single Level & Output \\
        U10     & 10m Wind Speed & m/s & Single Level & Output \\
        QFLX    & \textbf{Surface Moisture Flux} & m & Single Level & Output \\
        FSNS    & \textbf{Net Solar Flux} at Surface & J/m² & Single Level & Output \\
        FLNS    & \textbf{Net Longwave Flux} at Surface & J/m² & Single Level & Output \\
        FSNT    & \textbf{Net Solar Flux} at TOA & J/m² & Single Level & Output \\
        FLNT    & \textbf{Net Longwave Flux} at TOA & J/m² & Single Level & Output \\
        SHFLX   & \textbf{Sensible Heat Flux} & J/m² & Single Level & Output \\
        LHFLX   & \textbf{Latent Heat Flux} & J/m² & Single Level & Output \\
        \bottomrule
    \end{tabular}
\end{table}

\subsection{Training and Verification Data}

We use the Community Atmosphere Model version 6 (CAM6), the atmospheric component of the Community Earth System Model version 2.1.5 (CESM2), developed by the NSF NCAR \citep{bogenschutz2018path, gettelman2018projections}. At the training time, CAM6 was the latest supported model release and incorporates advancements in atmospheric physics, cloud microphysics, and boundary layer turbulence while leveraging a finite-volume (FV) dynamical core. 

For this study, we run CAM6 in AMIP mode, where it is forced by observed SSTs and sea ice concentrations from 1979 to 2014 \citep{huang2017noaa, huang2021improvements, rayner2003global}, with these monthly forcing fields linearly interpolated to daily values. CAMulator also accounts for time-evolving aerosol emissions and trace gas concentrations (including CO$_{2}$) to ensure consistency with historical atmospheric conditions.

Model simulations use the scientific release resolution of CAM6, with a horizontal grid spacing of 0.9$^\circ$ latitude $\times$ 1.25$^\circ$ longitude and 32 hybrid sigma-pressure levels extending up to 2.26 hPa in the vertical dimension. The emulator was trained on a number of static and dynamic variables (see Table \ref{tab:ml_variables}), all  saved at 6-hourly intervals. For simplicity, we keep the  vertical and horizontal resolution the same. 

Flux-form variables (see bold variables in Table \ref{tab:ml_variables}) and precipitation are treated as 6-hour accumulated quantities and rescaled so that downward fluxes are positive. Flux-form variables are essential for estimating sources and sinks in the atmospheric moisture and energy budgets. Precipitation (PRECT) represents the total precipitation leaving the column, including both parameterized and large-scale rain and snow. Prognostic variables are stored as 6-hourly averages. The prognostic variable Qtot represents the total water content in the model column, including vapor and condensed phases: specific humidity, as well as snow and rain that are present within the column.

Both static and dynamic forcing variables are included as input. Surface geopotential represents topography, while a land-sea mask distinguishes between ocean and land grid points to ensure accurate surface-atmosphere interactions. The model is also provided with dynamically forced (changing in time) variables, including sea surface conditions and incoming solar radiation, updated every 6 hours. The model predicts the next atmospheric state using only the previous time step (\( t \)), following an autoregressive formulation: \( t \rightarrow t+\Delta t \).

All variables, except for the Land-Sea Mask, undergo z-score normalization, where the mean and standard deviation are computed based on the training data period (1979–2010). The Land-Sea Mask represents the fraction of land within each grid cell, with values ranging between 0 and 1, and is left unnormalized. 

The goal of this work is to represent CAM6 with as much fidelity as possible, thus, we primarily analyze the emulation of CAM6 by CAMulator. In some cases, we incorporate reanalysis products to provide qualitative context for the differences between CAM6, CAMulator, and the assimilated observational datasets. For precipitation, we use NOAA’s Global Precipitation Climatology Project (GPCP) product \citep{Adler2003}, while all other variables are compared against the global ECMWF Reanalysis version 5 (ERA5) \citep{hersbach2020era5}. 

Following best practices, we divide the available data period into three subsets: a training data set (1979 - 2010), a validation subset (2011) and a testing data set (2012–2014).

\section{Methods}

\subsection{Model Architecture}

\begin{figure} 
    \centering
    \includegraphics[width=\columnwidth]{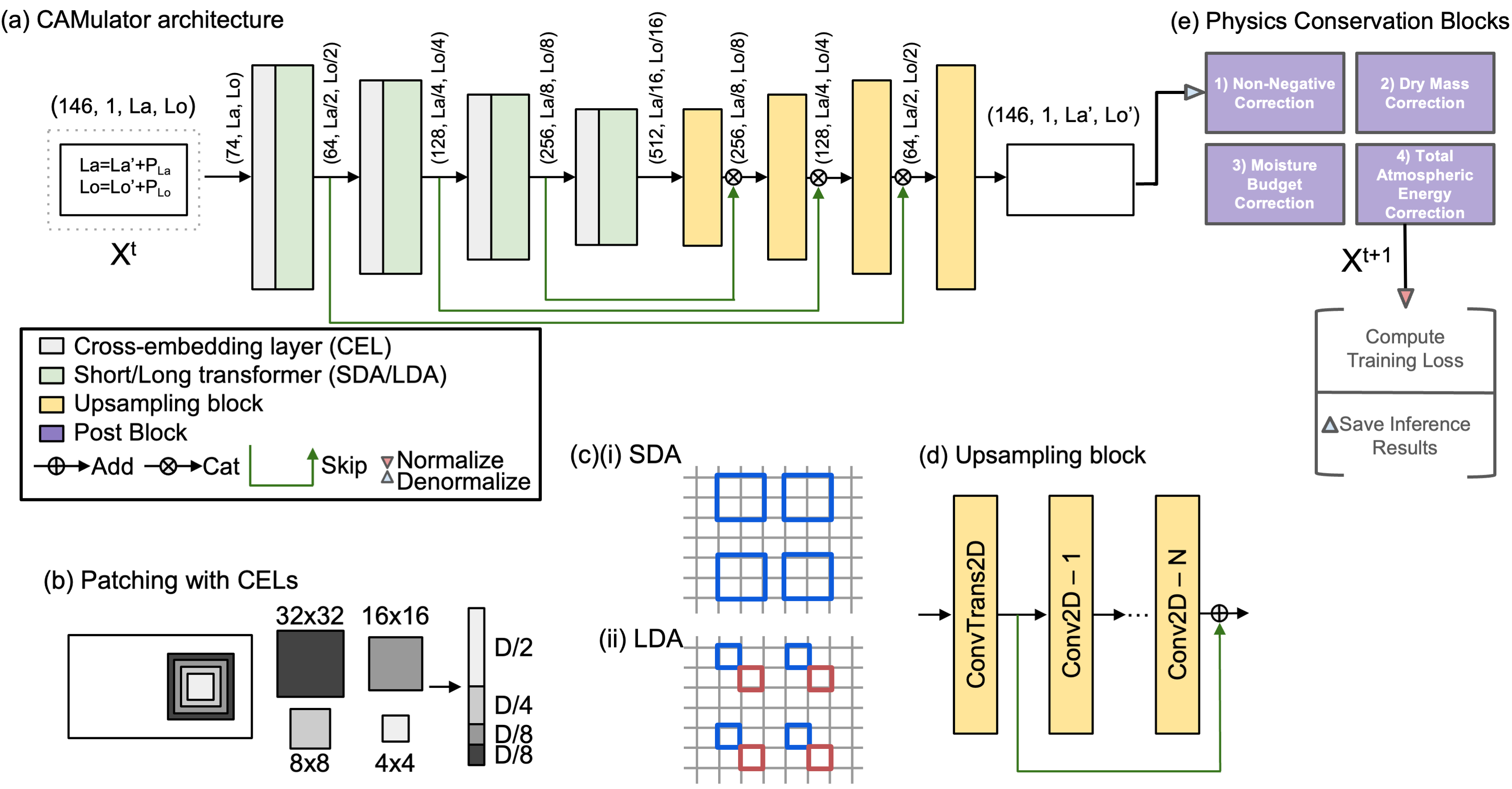}
    \caption{(a) the CAMulator architecture consists of encoding stages using a CrossFormer backbone and decoding stages with hierarchical transpose convolutional layers, with skip connections for improved feature flow. (b) The CEL captures multi-scale features using four convolutional kernels. The LSDA mechanism includes (c)(i) SDA for local interactions and (c)(ii) LDA for global dependencies. (d) The decoder component employs convolutional upsampling blocks with skip connections to progressively increase feature map resolution and maintain spatial information. e) predictions are then de-normalized and past through the four physics conservation blocks prior to loss calculations}
    \label{fig:CAM_schem}
\end{figure}

CAMulator is based on WXFormer, a transformer-based architecture developed at NSF NCAR and described in \cite{schreck2024community}. Figure \ref{fig:CAM_schem} shows the CAMulator architecture and workflow. WXFormer utilizes a CrossFormer backbone \citep{wang2023crossformer++} for multi-scale feature processing and long-range dependency modeling, combined with hierarchical transpose convolutional layers for upsampling in the decoding stages. Standard skip connections, similar to those in a U-Net \cite{ronneberger2015u}, are incorporated to efficiently transfer feature information and preserve spatial details from the encoder to the decoder. 

WXFormer has demonstrated state-of-the-art performance among leading AI-based weather prediction (AIWP) models \citep{schreck2024community}, making it an ideal base-model for climate modeling. To adapt WXFormer for our use case, we introduce several architectural modifications tailored to climate-scale forecasting.

First, we increase the feature embedding size by doubling the dimensionality of the cross embedding layers (CEL) while maintaining computational efficiency (Fig. \ref{fig:CAM_schem}b). The CEL employs a multi-kernel approach, applying four convolutional kernels (4×4, 8×8, 16×16, and 32×32) in parallel, each with a 2×2 stride during the initial processing stage \citep{wang2023crossformer++}. This multi-scale strategy enhances the model’s ability to capture spatial patterns across a range of atmospheric scales.

Additionally, we reduce both the global and local window sizes of the Long Short Distance Attention mechanism (LSDA; Fig. \ref{fig:CAM_schem}c) to align with the lower-resolution of CAM6 compared to WXFormer’s original $\sim$0.28° ERA5 application (see \citep{schreck2024community} for further details). This adaptation ensures more effective regional feature extraction while maintaining performance consistency.

With these modifications, CAMulator is comprised of \textasciitilde751 million trainable parameters, making it a high-capacity climate emulator optimized for long-term simulations. CAMulator takes as input the state of the atmosphere at time step $i$ and predicts the state at $i+1$ with a 6-hour forecast horizon. The model operates autoregressively, generating multi-step forecasts, allowing it to function as a standalone climate model capable of running indefinitely. 

\subsection{Conservation Blocks}

Following \citep{sha2025improving} and \cite{ watt2024ace2}, conservation schemes are applied after the CAMulator output layer to ensure physically consistent model roll-outs (Fig. \ref{fig:CAM_schem}e). The order of application is critical, as each scheme depends on the corrections applied in previous steps. Below, we describe the purpose and implementation of each scheme, with the adjusted variables emphasized. For the direct calculation of these corrections see Appendix section \ref{A01-cons}.

\begin{enumerate}
    \item \textbf{Nonnegative correction}: AI models can produce negative values, which are unphysical for certain variables. For all nonnegative variables (\textit{specific total water, total precipitation, 10-meter windspeed, and [Total, High, Low, Medium] Cloud Cover}), any negative raw outputs are set to zero. While this approach ensures physical consistency, alternative correction methods, such as redistribution, may be beneficial in some cases.
    
    \item \textbf{Global dry air mass conservation scheme}: At each forecast step, \textit{surface pressure} is corrected to ensure that global dry air mass remains constant, maintaining consistency with the initial condition. The correction is applied at each grid cell using a multiplicative factor, computed by first determining the residual mass imbalance and then adjusting accordingly.

    \item \textbf{Global moisture budget conservation scheme}: At each forecast step, \textit{total precipitation} is adjusted to balance the global sum of the total precipitable water tendency (derived from specific total water) and the accumulated net flux of precipitation and evaporation over the previous 6-hour period. The correction is applied at each grid cell using a multiplicative factor, computed by first determining the residual water imbalance and then adjusting accordingly.

    \item \textbf{Global total atmospheric energy conservation scheme}: The global atmospheric energy budget is defined as the balance between the tendency of total atmospheric energy and net energy fluxes at the top of the atmosphere and the surface. At each forecast step, \textit{Temperature} is corrected to ensure that the sum of total atmospheric energy tendencies aligns with the net energy sources and sinks over the past 6-hour period.

\end{enumerate}

Corrections to surface pressure, total precipitation, and air temperature are applied using multiplicative ratios across all grid cells, computed dynamically at each time step. While this approach enforces conservation, there are no explicit safeguards against overcorrection and no effort to redistribute values in the nonnegative correction, which may warrant further investigation in future work. We test the model in configurations with and without the physics blocks activated (CAMulator-phys and CAMulator-nophys, respectively).

It is essential to apply these conservation schemes in the specified order to preserve the theoretical dependencies between mass, moisture, and energy conservation. Further technical details are provided in Appendix \ref{A01-cons}.

\subsection{Training}
CAMulator was trained in a staged approach to balance stability and conservation constraints. Initially, the model was trained for 113 epochs as a single-step (6-hour) prediction task, minimizing the latitude-weighted mean squared error (MSE), with conservation block layers entirely omitted to allow for more stable initial learning. After this phase, conservation layers were introduced, and training continued for 77 additional epochs, now as a two-step (12-hour) forecast task. In the two-step training, the loss from both forecasted states was accumulated and used to optimize the model weights. For the single-step pretraining, cosine-annealing schedules were applied with an initial learning rate of $\mathrm{1E-5}$. Both stages used latitude-weighted MSE as a loss function, the AdamW optimizer \citep{loshchilov2017decoupled}, and batch sizes of 32. The training was conducted using Pytorch on 4 NVIDIA A100 GPUs, each with 40 GB of memory \citep{paszke2019pytorch}. In this phase final model state trained for 190 epochs in total over the course of three days. 

To determine the final model weights, we conducted a validation cycle, which consisted of training on 500 samples followed by a ten-year simulation. The resulting climatology was then compared to an archived CAM6 climatology. This process was repeated after each training iteration to assess the model’s ability to capture the climatology of precipitation and 2-meter temperature. The final model was selected based on latitude-weighted mean squared error (RMSE), ensuring the closest match to the CAM6 climatology for these two fields. The 50th cycle was selected as our final model, which took an additional 12 hours of training. 

\subsection{Climate Forecast Inference Options}

We explore multiple inference configurations, each defined by the forcing applied to the SST field, while all other dynamic forcing variables remain unchanged. The following four SST scenarios are considered:  

\begin{enumerate}
    \item \textbf{Observed SSTs (1979–2014)} – Uses monthly historical SST values from this period as in the CAM6 simulations
    \item \textbf{Year 2000 Climatological SST} – Applies the \textbf{mean SST state} from the year 2000, held constant throughout the simulation.
    \item \textbf{Year 2000 +2K Climatological SST} – Uses the \textbf{year 2000 SST climatology}, with a uniform and instantaneous \textbf{+2K temperature increase} applied globally to the SST field.
    \item \textbf{Year 2000 +4K Climatological SST} – Uses the \textbf{year 2000 SST climatology}, with a uniform and instantaneous \textbf{+4K temperature increase} applied globally to the SST field.
\end{enumerate}

For every SST case, the monthly sea surface values are linearly interpolated to a 6-hour timestep. In the 1979–2014 observed SST case, the simulation continuously cycles through these years, allowing for realistic variability. In contrast, climatological SST scenarios (year 2000, +2K, and +4K) can be extended indefinitely, making them well-suited for exploring equilibrium climate responses under different baseline ocean conditions. We note that the model was trained exclusively on the 1979–2010 SST state, making the remaining two inference cases (Year 2000+2K, and +4K climatological SSTs) entirely out-of-sample predictions. To ensure diverse initial conditions, we introduce a stochastic kinetic energy backscatter scheme \citep{berner2009spectral} for the first 15 days to the 1979–2014 SST runs, allowing the model to reach an independent atmospheric state by week 2. This approach mirrors the initialization strategy used in CAM6 simulations, where initial temperature perturbations help generate distinct trajectories \citep{kay2015community}.

\subsection{Computational Speed and Opportunities}\label{scomp}

CAM6 achieves significant computational throughput on Derecho, a NSF NCAR supercomputer, with 10 CPU nodes delivering 14 simulation years per day at the selected resolution. This performance metric reflects pure compute time, excluding I/O overhead. In contrast, our trained ML-based emulator achieves a dramatic 350 times speedup, running at 480 simulation years per day for a deterministic run on a single NVIDIA A100 GPU, while including all computational overhead and I/O operations. This acceleration enables high-throughput climate experiments, facilitating long-term scenario projections, ensemble simulations, and uncertainty quantification that would otherwise be computationally prohibitive with traditional numerical models.

While this speedup is already substantial, further improvements at inference time are possible through:
\begin{itemize}[noitemsep, topsep=0pt]
    \item \textbf{Scalability via Ensemble Parallelism}: Expanding ensemble members across multiple GPUs to improve robustness and uncertainty estimation.
    \item \textbf{Memory-Efficient Data Handling}: Leveraging \textbf{asynchronous I/O strategies} and distributed storage solutions to further reduce bottlenecks in large-scale climate simulations.
\end{itemize}

We avoid mixed precision (FP16/BF16) arithmetic to preserve the conservation properties critical for accurate climate modeling.

This optimized workflow positions the emulator as a scalable and efficient alternative to traditional GCMs, capable of running thousands of years of simulation within days, opening new avenues for large-ensemble climate forecasting, extreme event attribution, and policy-relevant decision support.

\section{Results}

\subsection{Annual Cycle and Roll-out}

\begin{figure} 
    \centering
    \includegraphics[width=\columnwidth]{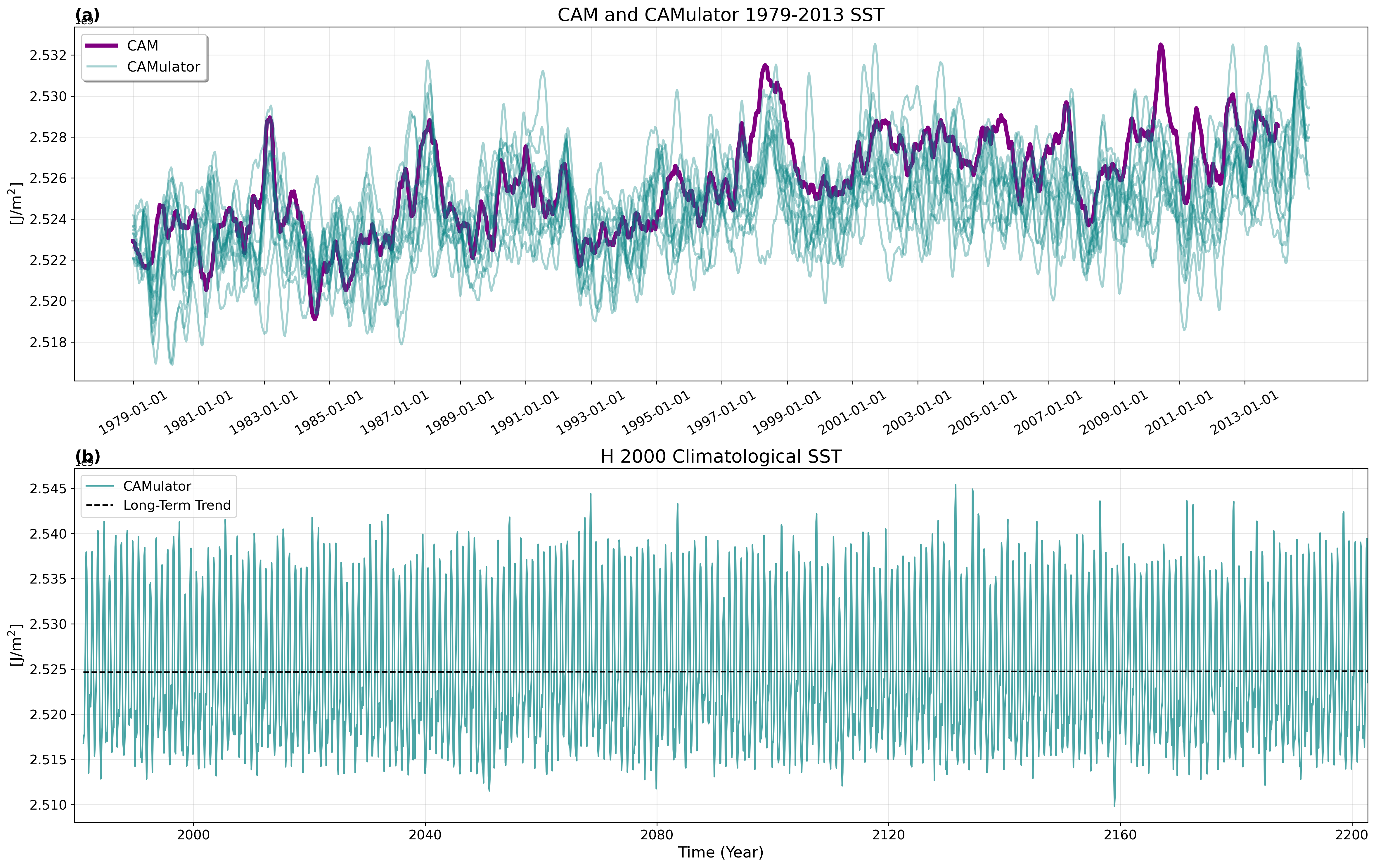}
    \caption{
    Column-integrated heat content in CAMulator to different SST forcing scenarios. (Top panel) A 12-member CAMulator ensemble (teal) is compared to the CAM6 simulation (purple) using observed SSTs from 1979–2014. CAMulator successfully captures the long-term warming trend and interannual variability. To isolate the trend, the seasonal cycle has been removed by regressing out the six leading harmonics, and a 90-day rolling mean has been applied. The ensemble spread arises from stochastic kinetic energy backscatter (SKEBS) perturbations applied during the first 15 days of simulation. (Bottom panel) CAMulator forced with fixed year-2000 climatological SSTs produces a stable long-term simulation with no discernible trend, demonstrating that the model does not introduce artificial warming in the absence of an external forcing mechanism.}
    \label{fig:TempTrend}
\end{figure}

Figure \ref{fig:TempTrend} illustrates two key aspects of CAMulator’s response to SST forcing. The top panel compares a 12-member CAMulator ensemble with the training data from 1979–2014, demonstrating that CAMulator effectively captures the long-term warming trend of total column-integrated heat content. Even though the emulator is initialized from weather states that then evolve dynamically at 6 hour time steps, it is still able to capture major modes of low-frequency internal variability while remaining on the correct trend trajectory. Thus, this serves as a semi-independent test of the emulator’s ability to reproduce climate-scale variability in a changing climate, despite the presence of the trend in the training data. To isolate this trend, the seasonal cycle was removed  by regressing out the six leading harmonics, and a 90-day rolling mean was applied to the time series (Fig. \ref{fig:TempTrend}a). The CAM6 simulation remains well within the ensemble spread of CAMulator, highlighting the ability of CAMulator to reproduce observed interannual variability. The CAMulator ensemble spread arises from the introduction of stochastic kinetic energy backscatter \citep[SKEBS, e.g.,][]{berner2009spectral}, which perturbs initial conditions over the first 15 days before allowing the system to evolve freely. We see an underestimation of total heat capacity in the period from 2003 to 2010 in CAMulator, however, CAM6 still largely sits within the ensemble spread. This discrepancy, is discussed further in ``Model Deficiencies and Future Improvements".

Figure \ref{fig:TempTrend}b demonstrates the model’s behavior when forced with fixed year-2000 climatological SSTs. In this scenario, CAMulator maintains an indefinite stable rollout with no emergent trend, as the absence of external forcing in SSTs prevents any sustained warming signal. This suggests that while the model responds effectively to imposed SST trends, it does not generate spurious warming in the absence of a forcing mechanism, and is capable of indefinite fixed climate rollouts. We show an identical figure to Figure \ref{fig:TempTrend} in the supplemental material, but for total water path (Fig. S1) and find similar results. 

Notably, across all SST forcing scenarios—including cases where the simulations were forced with SSTs outside of the training distribution—CAMulator has exhibited no signs of numerical instability. To date, no model crashes with our final model configuration have been observed, highlighting its robustness in handling a range of climate conditions.

\begin{figure} 
    \centering
    \includegraphics[width=\columnwidth]{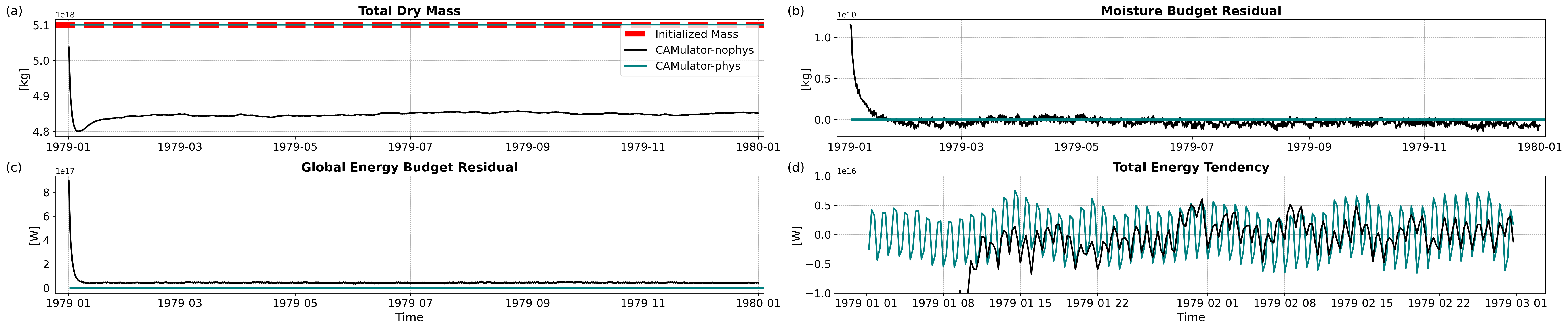}
    \caption{Time series of conservation diagnostics and budget residuals for CAMulator with and without physics conservation blocks. 
    (a) Total dry mass (kg) for CAMulator-phys (teal), CAMulator-nophys (black), with the initialized mass shown as a reference (red dashed line). 
    (b) Moisture budget residual (kg) comparing CAMulator-phys and CAMulator-nophys. 
    (c) Global energy budget residual (W) for CAMulator-phys and CAMulator-nophys. 
    (d) Total energy tendency (W), representing the time derivative of total atmospheric energy, comparing CAMulator-phys and CAMulator-nophys.}
    \label{fig:conservation}
\end{figure}

Figure \ref{fig:conservation} shows the conservation properties for global mass, water, and energy of the CAMulator system with the conservative layers activated (Figure \ref{fig:conservation}; CAMulator-phys teal line) and inactive (Figure \ref{fig:conservation}; CAMulator-nophys black line). CAMulator-nophys immediately deviates from the desired conservation properties and quickly settles into it's own errant climatology halfway through the first month of the climate simulation. The residuals are calculated as the previous time-step minus the current [$t_{-1} - t_0$], meaning an observed positive residual indicates that CAMulator-nophys sheds mass, water, and energy prior to settling into a steady state. We note that we can still achieve stable simulation runs and indefinite roll-outs with CAMulator-nophys. 

Interestingly, enforcing total energy tendency conservation balances the tendency such that it is centered around zero and regularly oscillates with the diurnal cycle (Fig. \ref{fig:conservation}e), whereas sporadic behavior is observed in CAMulator-nophys.

\subsection{Annual Mean Biases}

\begin{figure} 
    \centering
    \includegraphics[width=0.8\columnwidth]{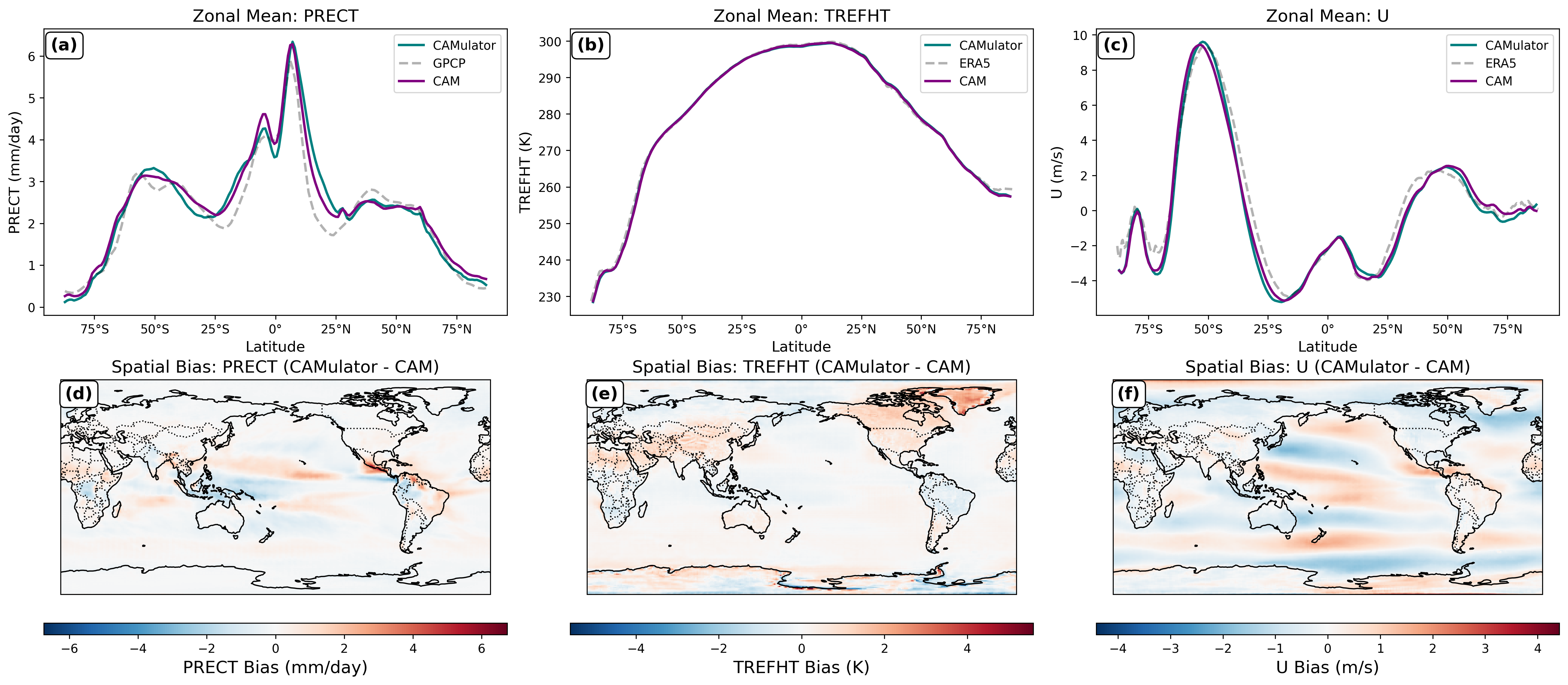}
    \caption{Zonal mean and spatial bias analysis of key climate variables. The top row (panels a–c) displays the zonal mean values for (a) precipitation rate (PRECT, mm/day), (b) near-surface air temperature (TREFHT, K), and (c) zonal wind component (U, m/s), comparing simulations from CAMulator (teal) and CAM6 (purple). Reanalysis products (GPCP [1981-2010] and ERA5 [1979-2010]) are shown shown in grey dash. The bottom row (panels d–f) presents the corresponding spatial biases (CAMulator - CAM) for (d) PRECT, (e) TREFHT, and (f) U, highlighting regional differences. Biases are computed as the annual mean differences between the two simulations and are visualized using a diverging colormap (red: CAMulator > CAM, blue: CAMulator < CAM)}
    \label{fig:prect_t2m_u_bias}
\end{figure}

We next evaluate the time-averaged annual climatology of the CAMulator simulation over the period 1979–2014. Figure \ref{fig:prect_t2m_u_bias}a–c shows the zonal mean precipitation, 2m temperature, and zonal wind at the lowest model level for CAM6 and CAMulator, shown in purple and teal, respectively. The zonal mean for two reanalysis products is also shown in dashed gray.

Overall, CAMulator simulates the annual mean state well. Figure \ref{fig:prect_t2m_u_bias}d–f presents the difference between the annual means (CAMulator - CAM). The largest precipitation errors occur in the tropics, with a wet bias over Central America and a dry bias over the Maritime Continent (Fig. \ref{fig:prect_t2m_u_bias}d). For 2m temperature, a persistent warm bias is evident over Greenland, particularly in winter, likely due to the lack of ice representation in the CAMulator state (Fig. \ref{fig:prect_t2m_u_bias}e). The largest discrepancies in zonal winds appear in the storm track regions, where slight shifts in the locations of peak wind magnitudes are observed (Fig. \ref{fig:prect_t2m_u_bias}f). For every field, the spatial annual climatological RMSE is computed and displayed in the supplemental material (Figs. \ref{fig:RMSE_by_Lev}-\ref{fig:LHFLX_RMSE}), we additionally compute the annual RMSE values which are in appendix table \ref{tab:variables_rmse}.

\subsection{Modes of Variability}

Evaluating a climate model's performance requires not only assessing its accuracy in representing climatological averages but also its ability to capture lower-frequency climate modes \citep[e.g.,][]{Phillips2014}. Given the vast number of modes of variability identified in the literature, a comprehensive analysis is impractical. Instead, we focus on three principal and well-documented modes used in major climate model evaluations: the Pacific North American Pattern (PNA), the North Atlantic Oscillation (NAO), and the El Niño–Southern Oscillation (ENSO) precipitation response. As these modes typically peak in boreal winter (defined here as December–February (DJF)), our analysis will center on their wintertime behavior. This section focuses on the representation of the PNA and NAO, while the following section will examine ENSO.

\begin{figure} 
    \centering
    \includegraphics[width=0.8\columnwidth]{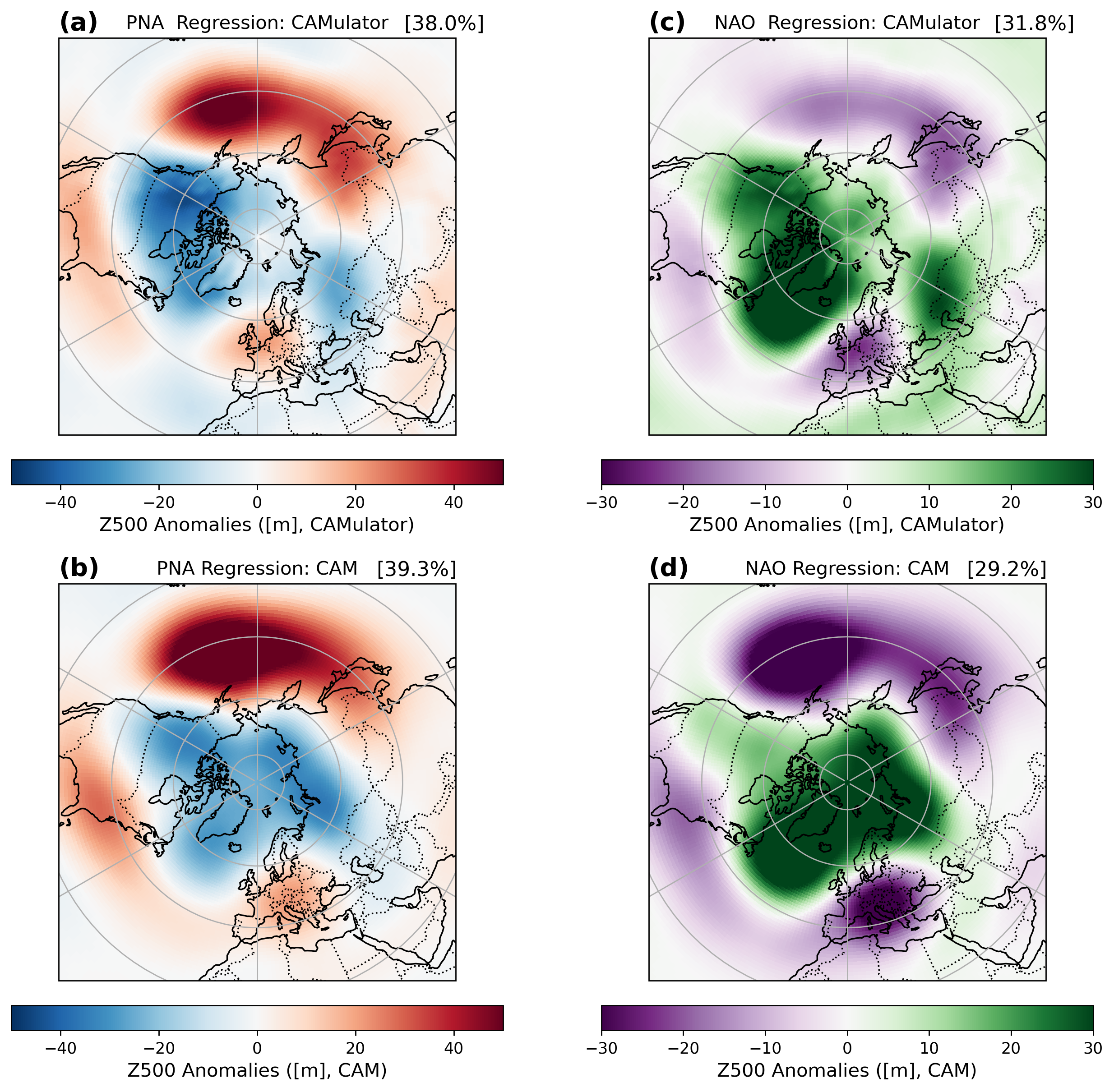}
    \caption{
    Regression of the anomalous Z500 field on the leading principal during DJF for the PNA region (left column, 20°–85°N, 120°E–120°W) and the NAO (right column, 20°–80°N, 90°W–40°E) over the years 1979–2014. Results are shown for CAMulator (top row) and CAM6 (bottom row). The explained variance of each mode is shown in the top right of each panel}
    \label{fig:PNA_NAO}
\end{figure}

Figure \ref{fig:PNA_NAO} shows the regression of the 500 hPa geopotential height (Z500) anomaly onto the leading principal component for the PNA and NAO regions in DJF, comparing CAMulator and CAM. Panels (a) and (b) illustrate the PNA-associated Z500 anomalies for CAMulator and CAM, respectively, while panels (c) and (d) display the NAO-related anomalies. Both patterns exhibit a close match to the CAM6 variability in terms of explained variance, indicating that these modes are well represented. Although the CAMulator patterns appear slightly muted in amplitude, they qualitatively capture the general structure of the CAM6 patterns and remain within the expected range of variability over a 30-year simulation \cite{simpson2020evaluation}. Notably, geopotential height is not directly predicted by CAMulator; rather, it is reconstructed by summing thicknesses across half-model levels to account for temperature and moisture variations between grid cells. The fact that CAMulator successfully reproduces the spatial structure of Z500 anomalies suggests that the physical coherence between surface pressure (PS), temperature (T), and total moisture (Qtot) has been well preserved, lending credibility to the model’s internal consistency.

\subsubsection{PNA \& NAO}

\subsubsection{ENSO Precipitation Response}

\begin{figure} 
    \centering
    \includegraphics[width=\columnwidth]{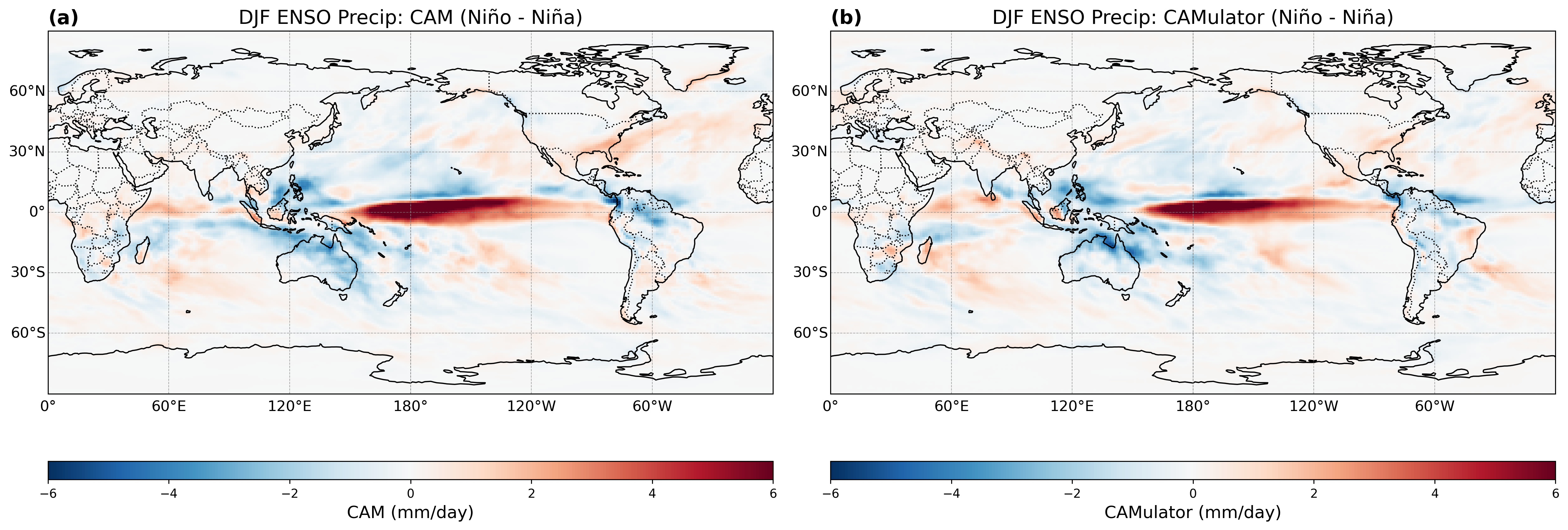}
    \caption{
    Composite difference in DJF precipitation (mm/day) during ENSO events (El Niño minus La Niña) for the eight strongest ENSO years on record, as identified by the Niño3.4 index. Results are shown for CAM6 (left) and CAMulator (right), with composites constructed using simple averaging. The Niño3.4-based El Niño years (December) include 1979, 1982, 1986, 1987, 1991, 1994, 1997, and 2002, while La Niña years (December) include 1983, 1984, 1988, 1995, 1998, 1999, 2000, and 2007. The color scale represents precipitation anomalies, where red indicates increased precipitation during El Niño relative to La Niña, and blue indicates reduced precipitation.}
    \label{fig:ENSO_Precip}
\end{figure}

ENSO is the dominant mode of tropospheric climate variability on interannual timescales, exerting a strong influence on global precipitation patterns \citep[e.g.,][]{philander1983nino, dai2000global, chapman2021monthly}. To assess how well CAMulator replicates the ENSO-related precipitation response, we analyze composite precipitation differences (El Niño minus La Niña) for the eight strongest ENSO events, identified based on absolute Niño3.4 SST anomalies over DJF and compare the response to CAM6 (Fig. \ref{fig:ENSO_Precip}a–b).

Both models capture the characteristic precipitation enhancements in the central and eastern Pacific, particularly near the International Date Line. However, CAMulator exhibits a slightly muted response in this region, a difference that may be attributable to internal variability. Notably, this muted response is consistent with the slight under representation of the PNA and NAO modes, suggesting a potential model tendency to underpredict low-frequency variability, which is notable in another ML-based climate model \cite{cresswell2024deep}. The precipitation derived from ENSO-driven teleconnections over North America are well captured, with drier conditions over the southern U.S. and wetter conditions in the Pacific Northwest. Similarly, the Maritime Continent response aligns closely between the models, indicating that CAMulator effectively represents the large-scale precipitation shifts associated with ENSO. Examining the patterns in Figure \ref{fig:ENSO_Precip} using a cosine latitude weighted Pearson correlation, we see a precipitation pattern correlation of 0.935 between CAM6 and CAMulator. Overall, CAMulator successfully reproduces the spatial patterns of the CAM6 ENSO precipitation response, showing strong qualitative agreement. 

\subsection{Extremes and Precipitation}

Beyond representing mean climate and large-scale modes of variability, a climate model’s ability to simulate extreme events is a crucial measure of its fidelity. Extremes (e.g. heatwaves, heavy precipitation events, or droughts) shape ecosystems, infrastructure resilience, and societal vulnerability. These high-impact phenomena arise from nonlinear interactions between atmospheric and oceanic processes, making their accurate representation a stringent test of a model’s physical consistency and predictive skill. Machine learning models, particularly those trained to minimize mean squared error, often struggle to capture extremes because they tend to favor the mean state over rare, high-magnitude events. However, CAMulator does not exhibit a strong tendency to underpredict extremes. We hypothesize that this may be due to its training methodology, which is limited to two prediction time steps (out to 12 hours), allowing it to better preserve variability. Supporting this, CAMulator’s kinetic energy and potential temperature spectra (see Supplemental Fig. \ref{fig:Spectra}) show minimal smoothing at smaller spatial scales, a common issue in data-driven models \citep[see,][]{rasp2024weatherbench} that can lead to weakened variability.

\begin{figure} 
    \centering
    \includegraphics[width=0.8\columnwidth]{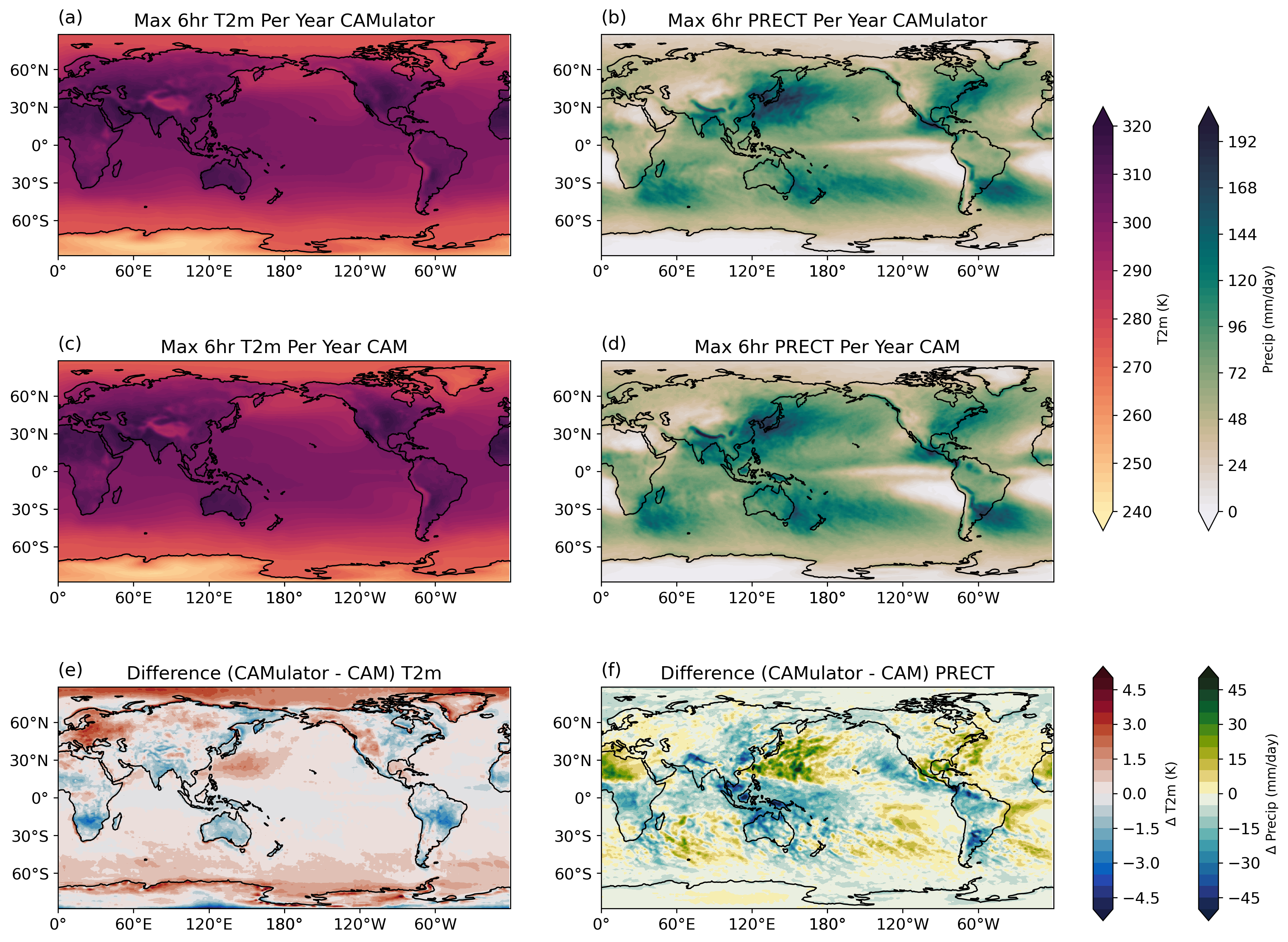}
    \caption{
    Climatology of the annual maximum 6-hourly average two-meter temperature (TREFHT, left column) and precipitation rate (PRECT, right column) over the period 1979–2014. The top row shows results from CAMulator, while the middle row presents results from CAM. The bottom row displays the differences (CAMulator - CAM), highlighting areas where CAMulator simulates warmer or colder extreme temperatures (red and blue shading, respectively) and higher or lower extreme precipitation rates (green and blue shading, respectively). Color bars indicate absolute values for TREFHT (K) and precipitation (mm/day) in the top and middle rows, while the bottom row represents their respective differences.}
    \label{fig:Extremes}
\end{figure}

Figure \ref{fig:Extremes} presents the annual maximum 6-hourly 2m temperature and precipitation at each grid point for CAM6 and CAMulator, with differences shown in the bottom row. Overall, CAMulator successfully captures the spatial patterns of extreme events, but some notable biases emerge. For 2m temperature, CAMulator overpredicts Arctic extreme temperatures, particularly in regions dominated by sea ice and land ice. This bias is expected, as CAMulator lacks explicit sea-ice and land-ice interactions—a limitation that will be addressed in future versions. For precipitation, CAMulator underpredicts extremes in the deep tropics while overpredicting them in the mid-latitudes. Notably, the region off the coast of Japan exhibits a strong positive bias in extreme precipitation, suggesting that CAMulator may be overestimating the intensity of tropical cyclones in this region.

\subsection{Rain Amount Distribution}

\begin{figure} 
    \centering
    \includegraphics[width=0.4\columnwidth]{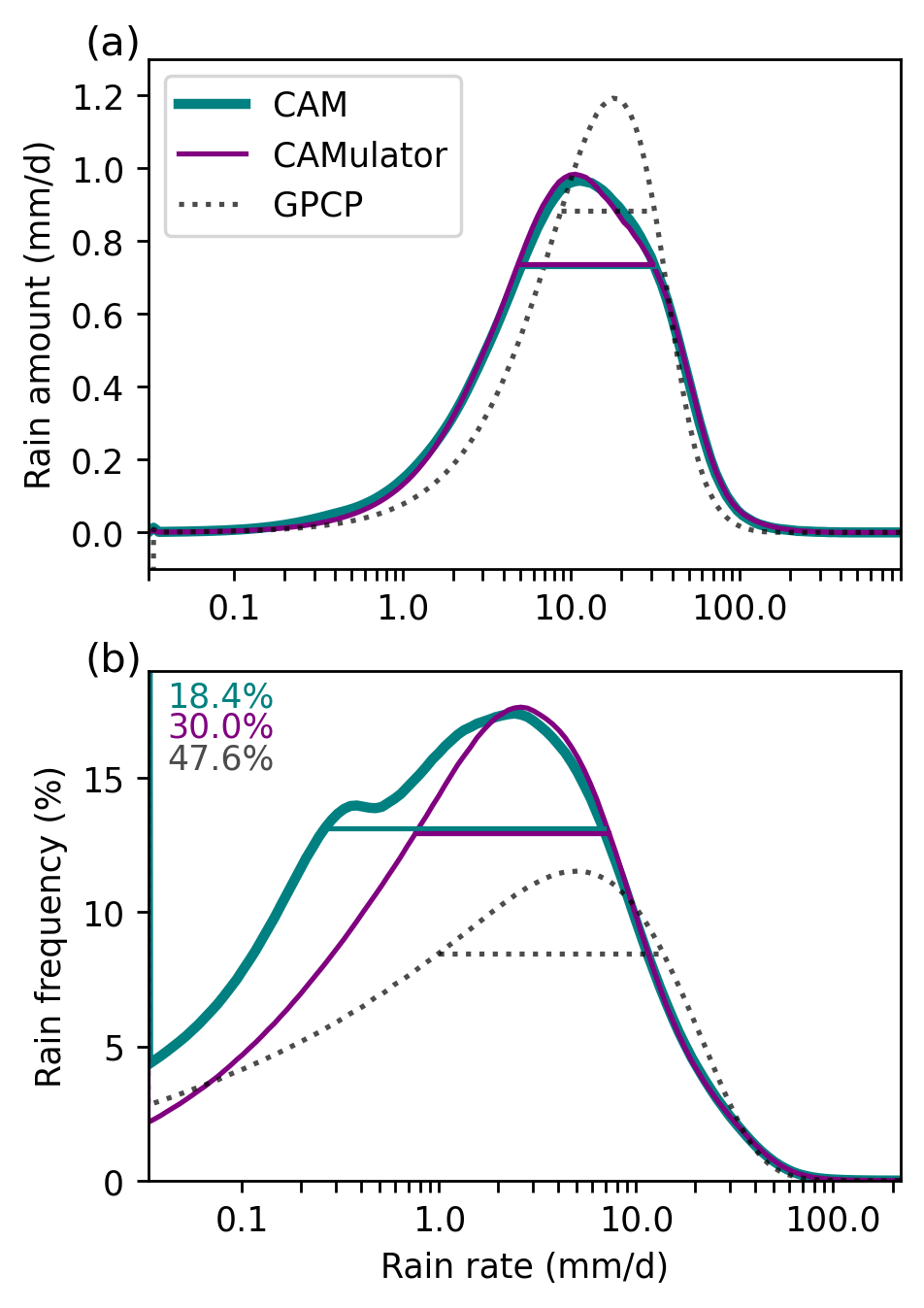}
    \caption{
    Global climatological distributions of daily rainfall from CAM6 (teal), CAMulator (purple), and GPCP observations (dotted black). 
    (Top) Rain amount (mm day$^{-1}$) as a function of rain rate. 
    (Bottom) Rain frequency distribution (\%) as a function of rain rate. 
    The dry-day frequency is indicated in the top left of the bottom panel, with values for CAM6 (teal), CAMulator (purple), and GPCP (black). 
    GPCP serves as an observational reference dataset and is coarsened prior to computing distributions.}
    \label{fig:rain_distribution}
\end{figure}

In this section we look at the similarities of the modeled rain amount and rain frequency distributions \citep[e.g.,][]{pendergrass2014two}, using logarithmically distributed rain-rate bins following \cite{watterson2003simulated}. See the appendix section \ref{A01} for the exact form of the calculation. The global distributions of CAM6 and CAMulator, along with the distribution observed in the daily GPCP dataset spanning years 1997-2012 are shown in Figure \ref{fig:rain_distribution}. The rain amounts in CAM6 and CAMulator are nearly identical, each peaking at around 10 mm day$^{-1}$ and are in agreement for all rain rates (Fig \ref{fig:rain_distribution}a). However, the biases in CAM6 when compared to the GPCP product (i.e. narrower distribution focused on a higher rain rate) persist in CAMulator. This is expected as CAMulator was trained as a CAM6 emulator. 

At larger rain rates (> 10 mm day$^{-1}$) the rain frequency distributions align fairly well (Fig \ref{fig:rain_distribution}b). However, at lighter rain rates, between 0.1 and 3 mm day$^{-1}$, there is a noted reduction in CAMulator rain frequency, bringing the distribution closer the the GPCP. Additionally, the dry-day frequency is nearly double that of CAM, and much more similar to GPCP. These are known biases in CAM6 that have persisted through most of the generations of the CAM models \citep[e.g.,][]{pendergrass2014two,pendergrass2017precipitation}. As discussed in \citep{sha2025improving}, the reduction in drizzle bias is primarily a consequence of our global conservation schemes. In our moisture budget framework, the global sum of total precipitation is required to be consistent with conserved quantities such as water, mass, and energy. When CAMulator overestimates drizzle—characterized by relatively low precipitation amounts—and underestimates precipitation in drier regions, the resulting global precipitation sum deviates from these conservation constraints. To enforce global consistency, a multiplicative correction is applied. However, because drizzle values are small, even minor absolute adjustments result in large relative changes. This disproportionate correction significantly increases the MSE during training, effectively penalizing the drizzle bias to meet the physical conservation laws.

\subsection{+2K and +4K runs}

\begin{figure} 
    \centering
    \includegraphics[width=\columnwidth]{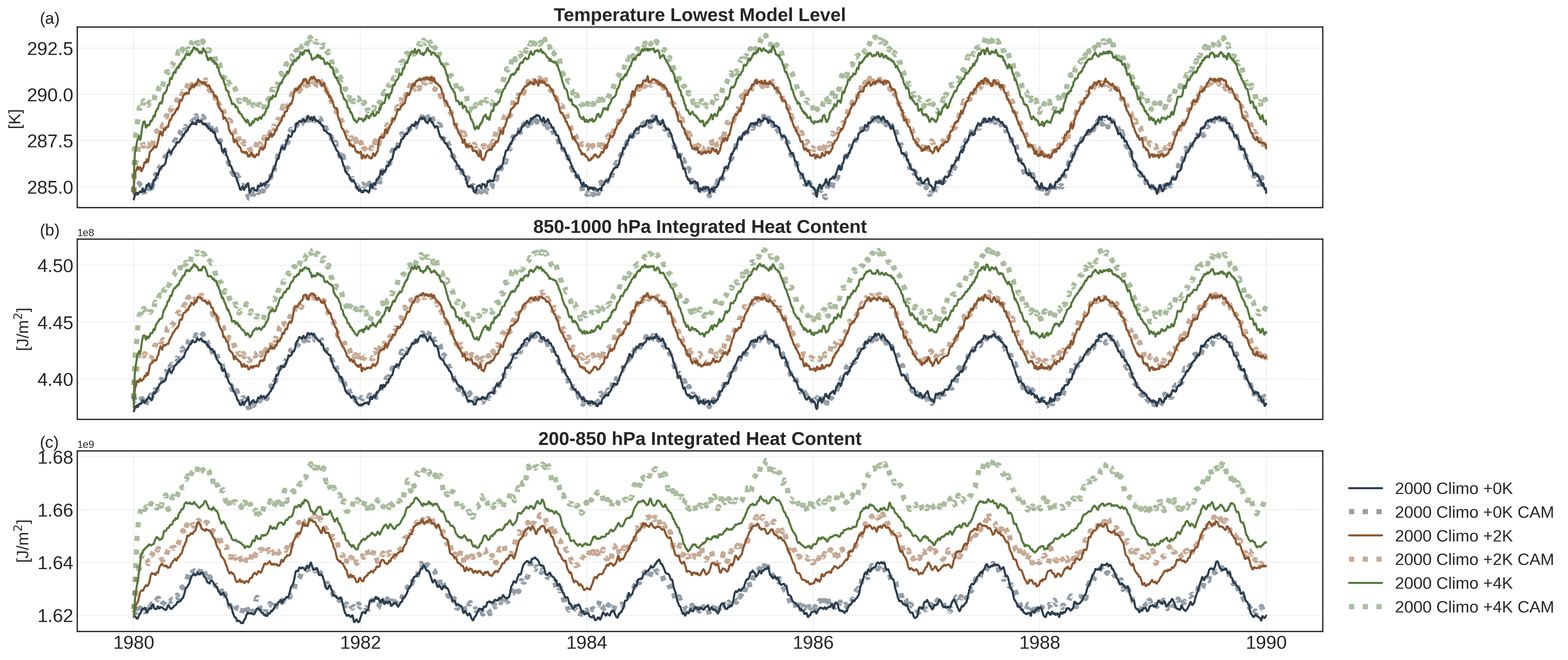}
    \caption{Time series of Global average Surface Temperature (a) and Column-integrated heat content (b,c) for simulations driven by different sea surface temperature (SST) climatologies: 2000 SST climatology (dark blue), 2000+2K (brown), and 2000+4K (green). The results from CAM6 (dashed lines) and CAMulator (solid lines) are shown. Integrated heat content is shown in two layers: (b) 850–1000 hPa and (c) 200–850 hPa.}
    \label{fig:TempTS2k4k}
\end{figure}

Before evaluating the CAMulator’s response to SST perturbations beyond its training distribution, we impose climatological SSTs for the year 2000 as a baseline forcing (+0K). Figure \ref{fig:TempTS2k4k} presents the global mean temperature at the lowest model level (Fig. \ref{fig:TempTS2k4k}a), along with the vertically integrated heat content for the lower (850–1000 hPa; Fig. \ref{fig:TempTS2k4k}b) and upper troposphere (200–850 hPa; Fig. \ref{fig:TempTS2k4k}c). To assess the model’s extrapolation capabilities under out-of-distribution forcing, we introduce uniform SST anomalies of +2K and +4K across all ocean grid cells and compare the CAMulator’s response to that of CAM6. Notably, these warming states lie well beyond the training data, testing the model’s ability to generalize under extreme forcing scenarios.

\begin{figure}[H]
    \centering
    \includegraphics[width=\columnwidth]{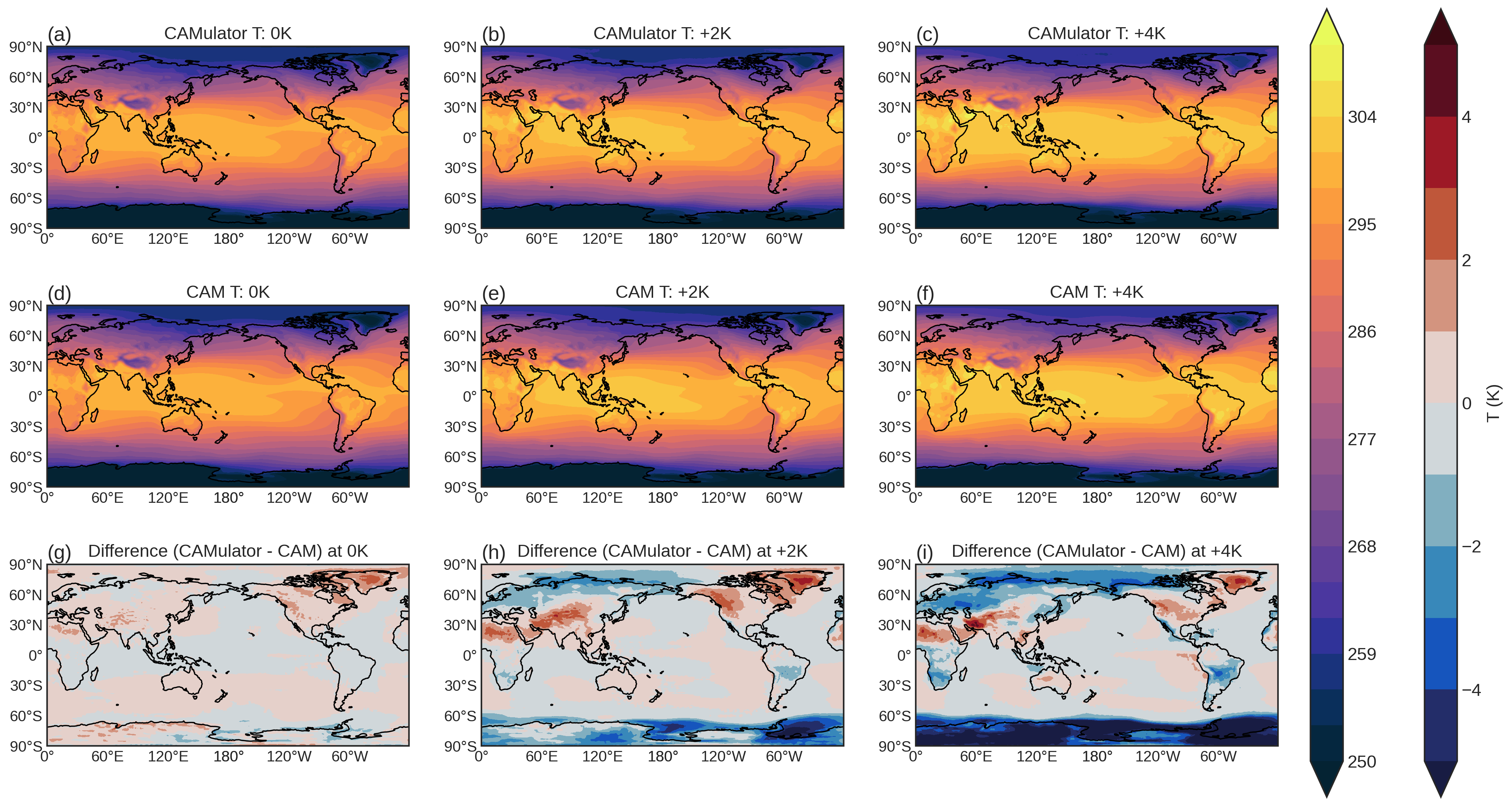}
    \caption{ 
    Annual mean temperature at the bottom model level in CAM6 (middle row) and CAMulator (top row) for three sea surface temperature (SST) climatologies: 2000+0K (left column), 2000+2K (middle column), and 2000+4K (right column). The bottom row shows the difference (CAMulator - CAM), highlighting deviations between the two models. The color scale represents temperature (K), with positive/negative values (red/blue) indicating stronger/weaker values in CAMulator.}
    \label{fig:TempSpat2k4k}
\end{figure}

The surface temperature response in the lowest model level exhibits the strongest agreement with CAM6, particularly in the year 2000 climatology +0K and +2K cases, where the seasonal cycles are well-aligned. However, discrepancies emerge under stronger warming, with a systematic cold bias in the CAMulator during boreal winter in both the +2K and +4K simulations. This bias suggests that CAMulator fails to fully capture key feedback mechanisms governing seasonal temperature variations in high-latitude regions. As shown in Figure \ref{fig:TempSpat2k4k}, the largest errors in 2-meter temperature occur in polar and sea-ice-dominated regions, indicating that the CAMulator lacks a robust representation of ice-atmosphere interactions. Specifically, the model has no explicit knowledge of sea-ice melt dynamics, surface albedo feedbacks, or key phase transition processes that modulate energy exchange in these environments. In contrast, CAM6 dynamically represents the retreat and expansion of ice cover, which strongly regulates the surface energy budget in these regions. The absence of such processes in the CAMulator likely leads to an unrealistic seasonal persistence of cold anomalies during boreal winter.

In the lower troposphere (Fig. \ref{fig:TempTS2k4k}b), the integrated heat content exhibits a seasonal amplitude shift relative to CAM6, which may reflect errors in boundary layer processes or the representation of heat fluxes from the surface. Similarly, upper-tropospheric heat content (Fig. \ref{fig:TempTS2k4k}c) shows systematic biases that may be linked to discrepancies in heating distribution via convective processes or the large-scale advection of heat anomalies.

Overall, the lack of cryosphere representation is a key limitation of CAMulator for extrapolation into warmer climates, particularly in regions where ice-related feedbacks play a critical role. Future work should assess whether incorporating explicit representations of land-ice/sea-ice state changes, or at least indirect constraints on polar energy fluxes, could mitigate these systematic biases and improve the model’s ability to handle out-of-distribution SST forcings.

\begin{figure} 
    \centering
    \includegraphics[width=\columnwidth]{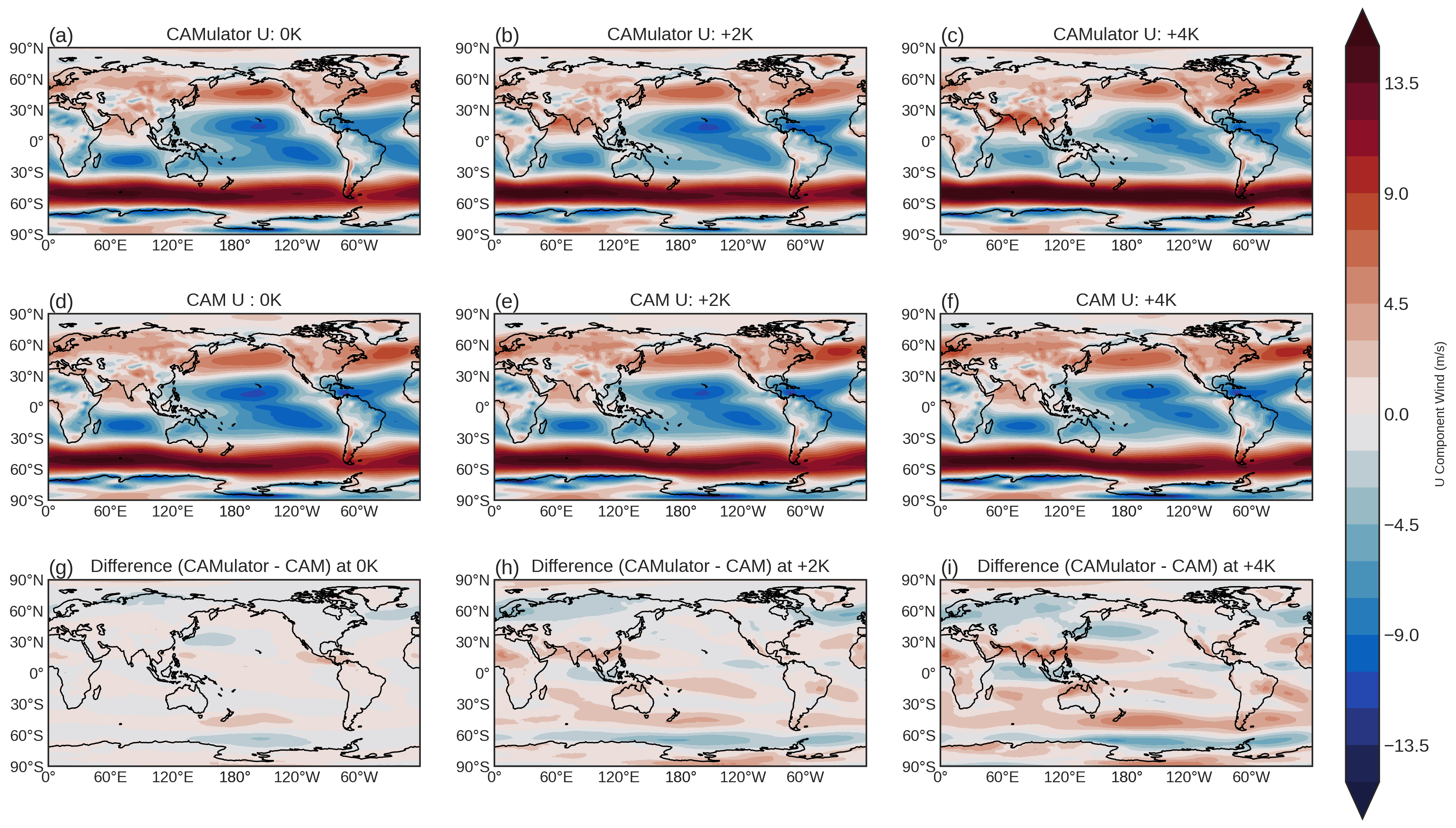}
    \caption{ 
    Annual mean zonal wind at the second model level in CAM6 (middle row) and CAMulator (top row) for three sea surface temperature (SST) climatologies: 2000+0K (left column), 2000+2K (middle column), and 2000+4K (right column). The bottom row shows the difference (CAMulator - CAM), highlighting deviations between the two models. The color scale represents zonal wind speed (m/s), with positive/negative values (red/blue) indicating stronger/weaker values in CAMulator.}
    \label{fig:Uwind2k4k}
\end{figure}

CAMulator’s ability to replicate large-scale low-level wind patterns under different SST forcing scenarios is evaluated in Figure \ref{fig:Uwind2k4k}. The top two rows show the zonal wind component ($U$) at 850 hPa for CAMulator (Figs. \ref{fig:Uwind2k4k}a–c) and CAM6 (Figs. \ref{fig:Uwind2k4k}d–f) across the +0K, +2K, and +4K SST warming states, while the bottom row (Figs. \ref{fig:Uwind2k4k}g–i) presents the difference between CAMulator and CAM6.

Overall, CAMulator demonstrates skill in capturing the large-scale structure of the low-level winds, particularly in the Southern Ocean. The core features of the midlatitude westerlies, including their equatorward contraction under warming, are well preserved. The close alignment with CAM6 suggests that CAMulator effectively learns the statistical relationship between SST forcing and the strength and position of the Southern Hemisphere storm tracks. Given that Southern Ocean winds are strongly tied to meridional temperature gradients and baroclinic wave activity and that this relationship is well represented in the training dataset, it is unsurprising that CAMulator generalizes well in this region.

On the other hand, the North Atlantic sector presents a key divergence from CAM6. Despite increasing SSTs, there is little evidence that the North Atlantic westerlies strengthen in response to warming in CAMulator, whereas CAM6 shows a more pronounced intensification. This discrepancy may reflect limitations in how CAMulator extrapolates the response of dynamically driven modes such as the North Atlantic Oscillation (NAO). Since the NAO is strongly influenced by internal atmospheric variability, rather than a direct SST-driven forcing, its response may be more difficult for a data-driven model to capture, especially in out-of-distribution scenarios. Additionally, CAMulator lacks an explicit representation of transient eddies and their feedbacks on mean flow, which could contribute to weaker NAO-like variability under warming.

The difference maps (Figs. \ref{fig:Uwind2k4k}g–i) further highlight systematic deviations, particularly in the North Atlantic and over regions of strong land-sea contrast, such as the western boundary currents. While these biases remain relatively small in magnitude compared to temperature biases, they suggest that CAMulator does not fully capture the dynamical adjustments that drive localized changes in low-level winds. This implies that some aspects of the wind field may be inherently more difficult for a ML-based emulator to reproduce from CAM6. 

Overall, these results indicate that CAMulator successfully replicates large-scale low-level wind features, particularly in regions where wind anomalies are tightly coupled to SST changes. However, for dynamically driven patterns such as the NAO, its response diverges from CAM6, potentially due to challenges in learning internally generated variability from training data alone. Future work should assess whether incorporating additional predictors—such as large-scale pressure anomalies or eddy kinetic energy—could improve CAMulator’s ability to mimic dynamically driven circulation responses to warming.

\section{CAMulator Deficiencies, Future Work, and Conclusions}

In this work, we introduce CAMulator, an auto-regressive, machine-learned (ML) emulator of the Community Atmosphere Model version 6 (CAM6). CAMulator is forced by sea surface temperatures (SSTs) and incoming solar radiation and is explicitly constrained to conserve global mass, water, and energy. It exhibits numerical stability over indefinite roll-outs and accurately reproduces the surface and integrated atmospheric heating response to within sample SST variations. The annual climatology is well captured, and dominant modes of variability, such as El Niño Southern Oscillation (ENSO), the North Atlantic Oscillation (NAO), and the Pacific North American (PNA) pattern, emerge naturally, with slightly muted amplitudes in some cases. Additionally, CAMulator’s physical constraints alleviate some of the drizzle problem commonly found in traditional climate models (Fig. \ref{fig:rain_distribution}). Beyond these physical attributes, CAMulator is computationally efficient and differentiable, making it a promising tool for a range of scientific applications.

\subsection{CAMulator Deficiencies and Future Improvements}

Despite these strengths, several key deficiencies remain, highlighting areas for future development.

\begin{itemize}
\item \textbf{High-latitude biases due to missing cryospheric processes:} The absence of interactive sea and land ice, and the limited representation of its change in the training period leads to a persistent cold bias, particularly in boreal winter. This bias is most pronounced in later periods and becomes apparent in seasonal temperature cycles (Supplemental Fig. \ref{fig:Annual_T2m_Cycle}). The lack of explicit ice-feedbacks and phase transition processes likely contributes to this discrepancy, especially in polar regions.

\item \textbf{Muted ENSO-related variability:}  While CAMulator successfully captures the broad-scale atmospheric response to ENSO events, it underestimates the magnitude of certain ENSO-related variability. This is particularly evident in the integrated atmospheric heat content during strong ENSO events, such as the 1997/98 El Niño (Fig. \ref{fig:TempTrend}). Improved representation of coupled ocean-atmosphere interactions and internal variability could enhance this response.

\item \textbf{Challenges in extrapolating to extreme SST perturbations:} CAMulator’s response to out-of-distribution SST forcing (+2K and +4K) diverges from CAM6, which suggests that while the model generalizes well within its training range, it struggles with conditions that require dynamically consistent extrapolation. Future improvements to training strategies, including exposure to a broader range of climate states and variables, may mitigate this issue.
\end{itemize}

To address these limitations, several promising research directions should be explored:

\begin{itemize}
    \item \textbf{Coupling CAMulator with an interactive ocean, sea ice, and land model:} Incorporating dynamic surface processes would improve the representation of feedback mechanisms, particularly in polar regions, and help resolve biases associated with missing feedback states. We hope that these models could be both physics-based and machine learned \citep[e.g.,][]{dheeshjith2024samudra}.

    \item \textbf{Enhancing variability through stochastic parameterizations:} \textit{Training} CAMulator with a Stochastic Kinetic Energy Backscatter Scheme \citep[SKEBS;][]{berner2009spectral} or a similar approach could improve CAMulator’s ability to represent subgrid-scale variability, potentially addressing the muted ENSO and NAO responses.

    \item \textbf{Exploring supermodeling approaches:} One promising avenue is to integrate CAMulator with multiple ML-based emulators coupled to a single dynamical model, allowing for dynamic state corrections and improved climate change projections \citep[e.g.,][]{schevenhoven2023supermodeling}. Such a framework could enhance the representation of future warming scenarios while maintaining computational efficiency.
\end{itemize}

\subsection{Broader Implications and Future Outlook}

The ability of CAMulator to rapidly emulate atmospheric states while conserving fundamental physical properties makes it an attractive tool for accelerating climate simulations, uncertainty quantification, and data assimilation. Furthermore, its differentiability opens new opportunities for inverse modeling and sensitivity analyses, potentially improving parameter estimation in Earth system models.

As ML-driven climate modeling continues to evolve, hybrid approaches that combine data-driven methods with traditional physics-based models will likely become increasingly valuable. CAMulator provides a foundation for such efforts, demonstrating that machine-learned emulators can maintain physical fidelity while offering substantial computational advantages. By incorporating additional coupling mechanisms and improving variability representation, future iterations of CAMulator could play a critical role in next-generation climate modeling, bridging the gap between computational efficiency and physical realism.

\section*{Acknowledgments}
This project is supported by Schmidt Sciences, LLC. We thank all the scientists and administrators who contributed to the development of CESM2. Additionally, this work was supported by the U.S. National Science Foundation National Center for Atmospheric Research (NSF NCAR), which is a major facility sponsored by the NSF under Cooperative Agreement No. 1852977 and by Grant No. RISE-2019758. Any opinions, findings, and conclusions or recommendations expressed in this material are those of the authors and do not necessarily reflect the views of the U.S. National Science Foundation. 

\section*{Open Science}

All figures in this manuscript can be reproduced using the scripts and notebooks available in the dedicated figure repository: \href{https://github.com/WillyChap/CAMulator_Figures}{CAMulator Figures GitHub}. 

The CAMulator model itself is developed on the NSF NCAR CREDIT platform \citep{schreck2024community} and is actively maintained. The specific branch used for climate simulations can be accessed here: \href{https://github.com/NCAR/miles-credit/tree/CAMulator_V01.00}{NCAR CREDIT Climate Runs Branch}.

\appendix
\section*{Appendix Material} % Custom title
\section{Metric Calculations} % Custom title

\subsection{Climatological RMSE}\label{A00}

The weighted annual climatological root mean squared error is calculated as:
\begin{equation}
\mathrm{RMSE}(t_i, t_l) = \sqrt{\frac{1}{N_{\phi} N_{\lambda}} \sum_{i=1}^{N_{\phi}}{ \sum_{j=1}^{N_{\lambda}} {\left\{w(i) \left[ \overline{CAM6(i, j)} - \overline{CAMulator(i, j)}\right]^2\right\}}}}
\end{equation}

\noindent
Where $N_{\phi}$ and $N_{\lambda}$ are the number of latitude ($\phi$) and longitude ($\lambda$) grid cells, respectively. $w\left(i\right) = \cos(\phi_i)$ is the latitude weighting coefficient, which is normalized such that \( \sum_{i} w(i) = 1 \), the over bar indicates that each field was first averaged in time over the 30 year simulation run. Table \ref{tab:variables_rmse} shows the RMSE calculated for each single level variable. 

\begin{table}[H]
    \centering
    \begin{tabular}{lllc}
        \toprule
        \textbf{Short Name} & \textbf{Long Name} & \textbf{Units} & \textbf{Globally Averaged RMSE} \\
        \midrule
        PRECT   & Precipitation Rate & mm/day & 0.5766 \\
        CLDTOT  & Total Cloud Cover  & fraction     & 0.0402 \\
        CLDHGH  & High Cloud Cover   & fraction     & 0.0391 \\
        CLDLOW  & Low Cloud Cover    & fraction     & 0.0340 \\
        CLDMED  & Medium Cloud Cover & fraction     & 0.0296 \\
        TAUX    & Zonal Wind Stress  & N/m²   & 0.0110 \\
        TAUY    & Meridional Wind Stress & N/m² & 0.0055 \\
        U10     & 10m Wind Speed     & m/s    & 0.3062 \\
        QFLX    & Surface Moisture Flux & kg/m²/s & 0.0474 \\
        FSNS    & Net Solar Flux at Surface & W/m² & 6.4269 \\
        FLNS    & Net Longwave Flux at Surface & W/m² & 2.6681 \\
        FSNT    & Net Solar Flux at TOA & W/m² & 5.8499 \\
        FLNT    & Net Longwave Flux at TOA & W/m² & 3.9247 \\
        SHFLX   & Sensible Heat Flux & W/m² & 2.1538 \\
        LHFLX   & Latent Heat Flux   & W/m² & 5.4987 \\
        \bottomrule
    \end{tabular}
    \label{tab:variables_rmse}
    \caption{Spatial Root Mean Square Errors (RMSE) for climate variables, including their names, descriptions, and measurement units. The RMSE values represent the deviation of CAMulator's 30-year climatological mean from the climatology of the base CAM6 simulation.}
\end{table}

\subsection{Atmospheric Heat Capacity}\label{A02}

The atmospheric heat capacity is calculated by integrated temperature in the vertical and taking a cosine-latitude weighted sum. 

\begin{equation}
    \frac{Cp}{g} \int_{P_1}^{P_0} T \, dP
\end{equation}

Where $Cp$ is the dry air specific heat capacity and is set to 1004 $[J kg^{-1} K]^{-1}$

\subsection{Rain amount and Rain Frequency}\label{A01}

The calculation of rain amount and rain frequency follows the method outlined in \cite{pendergrass2014two}, but we summarize the approach here for completeness. The rain-rate distribution is constructed using logarithmically spaced bins, where each bin is 7\% wider than the previous one, with its center shifted accordingly. Only rain rates exceeding 0.03 mm day\(^{-1}\) are included, while dry days are defined by a precipitation threshold of 0.0321 mm day\(^{-1}\).  

At each grid point, we first compute a histogram of rain rates and normalize it by the total number of days to obtain the rain frequency distribution. The total precipitation amount within each bin is then summed to construct the rain amount distribution. Finally, these distributions are averaged globally using an area-weighted approach to produce the global-mean distributions.  
To formalize this, we define the probability distribution of rain amount \( p \) and rain frequency \( f \) for each dataset, using daily rain accumulation \( r \) from model output or gridded observations. The bin edges are denoted as \(  T_i^l \) (left) and \(  T_i^r \) (right), with the bin centers defined as \(  T_i^c = ( T_i^l +  T_i^r)/2 \). The transformation of the distribution is expressed mathematically as follows:  

\begin{equation}
    p_i( T_i^c) = \frac{1}{\Delta \ln R} \int_{\ln  T_i^l}^{\ln  T_i^r} p(\ln r) d\ln r
    = \frac{1}{\Delta \ln T} \sum_{\text{gridpts}} r( T_i^l < r <  T_i^r) \frac{A_{\text{gridpt}}}{A_{\text{total}}},
    \label{eq:rain_amount}
\end{equation}

\begin{equation}
    f_i( T_i^c) = \frac{1}{\Delta \ln T} \int_{\ln  T_i^l}^{\ln  T_i^r} f(\ln r) d\ln r
    = \frac{1}{\Delta \ln T} \sum_{\text{gridpts}} \frac{N_d( T_i^l < r <  T_i^r)}{\sum N_d} \frac{A_{\text{gridpt}}}{A_{\text{total}}},
    \label{eq:rain_frequency}
\end{equation}

\begin{equation}
    F_d = \frac{1}{\sum N_d} \sum_{\text{gridpts}} N_d(r = 0) \frac{A_{\text{gridpt}}}{A_{\text{total}}},
    \label{eq:dry_days}
\end{equation}

where \( A \) is the grid cell area, and \( N_d \) represents the number of days in the dataset. The bin width is set as  

\begin{equation}
    \Delta \ln T = \frac{(T_{i+1} -  T_i)}{ T_i} = 7.67\%,
\end{equation}  

ensuring sufficient resolution to capture the distribution of rain rates.  

\subsection{Low Frequency Variability}\label{A01-MOV}

To extract the leading patterns of variability, we perform an empirical orthogonal function (EOF) decomposition on the monthly anomaly fields. The climatology is defined as the monthly mean computed over the full 30-year simulation or reanalysis record. Prior to decomposition, all fields are area-weighted by the square root of the cosine of latitude to account for variations in grid cell area. We represent the spatial patterns by regressing the field pointwise onto a one-standard-deviation change in the corresponding temporal principal component.

We examine the DJF Pacific–North American (PNA) pattern and the North Atlantic Oscillation (NAO) in detail. Following \citep{Phillips2014} (NCAR’s Climate Variability Diagnostic Package), these patterns are defined as the leading modes of atmospheric variability over the regions [20–85°N, 120°E–120°W] and [20–80°N, 90°W–40°E], respectively.

The eight strongest El Niño and La Niña events are identified using the Niño3.4 index from the observational record. The December El Niño years include 1979, 1982, 1986, 1987, 1991, 1994, 1997, and 2002, while the December La Niña years include 1983, 1984, 1988, 1995, 1998, 1999, 2000, and 2007. The composite precipitation pattern is then calculated by computing the DJF seasonal averages and subtracting the La Niña composite field from the El Niño composite field.

\section{Model Grid and Conservation Schemes}\label{A01-cons}

CAMulator leverages a hybrid sigma-pressure coordinate system, where vertical integrals are computed by first determining the local pressure value. The pressure at each model level \( k \) is given by:

\begin{equation}
P(i, j, k) = A_k P_0 + B_k PS(i,j)
\end{equation}

where \( P(i, j, k) \) represents the pressure at level \( k \) for a given latitude-longitude point \( (i, j) \). The terms \( A_k \) (hPa) and \( B_k \) (dimensionless fraction) are reference coefficients defining the hybrid coordinate system, while \( P_0 \) (1000.0 hPa) is the reference pressure. The surface pressure, \( PS(i,j) \), varies across grid points and is used to determine the pressure levels dynamically.

For a quantity $S(z)$ that varies with height $z$, its mass-weighted vertical integral can be converted to a pressure level integral using the hydrostatic equation:

\begin{equation}\label{sec2_eq3}
\int_{0}^{\infty}{\rho S}dz = \frac{1}{g}\int_{p_s}^{0}Sdp \approx \frac{1}{g}\int_{p_s}^{p_0}Sdp
\end{equation}

where \( S \) represents the variable of interest at the midpoint of the \( (i,j,k) \)-the grid box, and \( g \) is the acceleration due to gravity. We conduct similar corrections to \citep{sha2025improving,sha2025investigating}, except applied on sigma-hybrid pressure levels, we repeat the derivation of those calculations below for completeness.

\subsection{Global dry air mass conservation}\label{A12}

The evolution of dry air mass within a given atmospheric column is determined by the divergence of the vertically integrated dry air mass flux:

\begin{equation}\label{sec2_eq4}
\frac{1}{g}\frac{\partial}{\partial t}\int_{p_1}^{p_0}{\left(1-q\right)}dp = -\mathbf{\nabla} \cdot \frac{1}{g} \int_{p_1}^{p_0}{\left[\left(1-q\right)\mathbf{v}\right]}dp
\end{equation}

\noindent
where $\mathbf{v}$ represents velocity, and $q$ corresponds to total atmospheric moisture, approximated by specific total water (see Table \ref{tab:ml_variables}).

For a global sum, if the atmosphere is considered incompressible, the divergence term in equation (\ref{sec2_eq4}) becomes zero. Consequently, the total global dry air mass ($\langle M_d \rangle$) remains conserved over time (henceforth, $\langle \rangle$ denotes a global sum):

\begin{equation}\label{sec2_eq5}
\frac{\partial}{\partial t} \langle M_d \rangle = \frac{\partial}{\partial t} \left\langle \frac{1}{g}\int_{p_1}^{p_0}{\left(1-q\right)}dp\right\rangle = \epsilon_d
\end{equation}

\noindent
Where $\epsilon_d$ is the residual term that violates the global dry air mass conservation.

For two forecast time steps separated by $\Delta t = t_1 - t_0$, where $t_0$ corresponds to the initial analyzed state and $t_1$ denotes a subsequent validation time, equation (\ref{sec2_eq5}) can be reformulated as:

\begin{equation}\label{sec2_eq6}
\frac{\partial}{\partial t} \langle M_d \rangle =\langle M_d \left(t_0\right)\rangle - \langle M_d\left(t_1\right)\rangle = \epsilon_d
\end{equation}

During the correction stage, $PS$ can be adjusted to ensure $\epsilon_d=0$ using a multiplicative ratio. For this correction, the contribution of global dry air mass from coefficients $A$ and $B$ are estimated as follows:

\begin{equation}\label{sec2_eq5_b}
\begin{aligned}
\langle M_A \rangle = \text{SUM}\left[\frac{1}{g}\sum_{i_l=0}^{N_l-1}{\Delta A_{i_l}\left(1-q\right)_{i_l}}\right],\quad \Delta A_{i_l} = A_{i_l} - A_{i_l-1}\\[6pt] 
\langle M_B \rangle = \text{SUM}\left[\frac{p_s}{g}\sum_{i_l=0}^{N_l-1}{\Delta B_{i_l}\left(1-q\right)_{i_l}}\right],\quad \Delta B{i_l} = B_{i_l} - B_{i_l-1}
\end{aligned}
\end{equation}

Where $\langle M_A \rangle$ and $\langle M_B \rangle$ are global dry air mass components spread to $A$ and $B$, respectively. When computed on $t_1$, they are denoted as $\langle M_A \left(t_1\right)\rangle$ and  $\langle M_B \left(t_1\right)\rangle$.

The correction of $p_s$ is defined as follows:

\begin{equation}\label{sec2_eq6_b}
p_s^*\left(t_1\right) = p_s\left(t_0\right)\frac{\langle M_d \left(t_1\right)\rangle - \langle M_A \left(t_1\right)\rangle}{\langle M_B \left(t_1\right)\rangle}
\end{equation}

\noindent
Where $\langle M_d \left(t_0\right)\rangle$ is the total amount of global dry air mass calculated from the initial condition. $p_s^*\left(t_1\right)$ is the corrected $p_s$ on $t_1$. The same multiplicative correction is applied to $p_s$ on all grid cells.

\subsection{Global moisture budget conservation scheme}\label{A13}

For an air column, the time tendency of total precipitable water ($M_v$) is governed by the balance between the vertically integrated moisture flux divergence, evaporation, and total precipitation:

\begin{equation}\label{sec2_eq8}
\frac{\partial}{\partial t}M_v = \frac{1}{g}\frac{\partial}{\partial t}\int_{p_1}^{p_0}{q}dp = -\mathbf{\nabla} \cdot \frac{1}{g} \int_{p_1}^{p_0}{\left(\mathbf{v}q\right)}dp - E - P
\end{equation}

\noindent
Here, $q$ represents the specific total water content, while $E$ and $P$ correspond to evaporation and total precipitation, respectively, with units of $\mathrm{kg\cdot m^{-2} \cdot s^{-1}}$.

On a global scale, the divergence term in equation (\ref{sec2_eq8}) vanishes, implying that the global sum of $M_v$ is primarily modulated by the spatially averaged evaporation ($\langle E \rangle$) and precipitation ($\langle P \rangle$), with an additional residual term $\epsilon_m$ indicating conservation errors:

\begin{equation}\label{sec2_eq9}
-\left\langle\frac{\partial M_v}{\partial t}\right\rangle - \langle E  \rangle - \langle P \rangle = \epsilon_m
\end{equation}

\noindent
By convention, downward precipitation results in a positive $\langle P \rangle$, whereas $\langle E \rangle$ is typically negative due to upward evaporation.

To enforce moisture budget closure, precipitation is adjusted during the correction step via a multiplicative factor:

\begin{equation}\label{sec2_eq10}
P^*\left(t_1\right) = P\left(t_1\right)\frac{\langle P^*\left(t_1\right)\rangle}{\langle P\left(t_1\right)\rangle}, \quad\langle P^*\left(t_1\right)\rangle = -\left\langle\frac{M_v\left(t_1\right) - M_v\left(t_0\right)}{\Delta t}\right\rangle - \langle E\left(t_1\right)\rangle
\end{equation}

\noindent
where $\langle P^*\left(t_1\right)\rangle$ denotes the globally adjusted precipitation necessary to achieve moisture conservation. This correction is uniformly applied across all grid points.

% --------------------------------------------------------------------------------------------------------- %
\subsection{Global total atmospheric energy conservation scheme}\label{A14}

For a given air column, its vertically integrated total atmospheric energy ($A$) is defined as follows:

\begin{equation}\label{sec2_eq11}
A = \frac{1}{g}\int_{p_1}^{p_0}{\left(C_pT+L_v q+\Phi_s+k\right)}dp
\end{equation}

\noindent
The terms on the right side of the equation (\ref{sec2_eq11}) are thermal energy, latent heat energy, potential energy, and kinetic energy, respectively. $L_v$ is the latent heat of vaporization, and $\Phi_s$ is the geopotential at the surface. Kinetic energy ($k$) is defined as $k=0.5\left(\mathbf{v} \cdot \mathbf{v}\right)$. The specific heat capacity of air at constant pressure ($C_p$) is defined as $C_p=C_{pd}(1-q)+C_{pv}q$. The formulation of equation (\ref{sec2_eq11}) has some limitations, which will be addressed in a separated section below.

The tendency of $A$ is determined by the divergence of vertically integrated moist static energy ($h=C_pT+L_v q+\Phi$), kinetic energy, and other energy sources and sinks:

\begin{equation}\label{sec2_eq12}
\frac{\partial}{\partial t}A = -\mathbf{\nabla}\cdot\frac{1}{g}\int_{p_1}^{p_0}{\mathbf{v}\left(h+k\right)}dp = R_T - F_S
\end{equation}

\noindent
Where $R_T$ and $F_S$ are net radiation and energy fluxes on the top of the atmosphere and the surface.

\begin{equation}\label{sec2_eq13}
\begin{aligned}
R_T &= \mathrm{TOA}_{\mathrm{net}} + \mathrm{OLR} \\
F_S &= R_{\mathrm{short}} + R_{\mathrm{long}} + H_s + H_l
\end{aligned}
\end{equation}

\noindent
Where $\mathrm{TOA}_{\mathrm{net}}$ is the top-of-atmosphere net solar radiation, $\mathrm{OLR}$ is outgoing longwave radiation. $R_{\mathrm{short}}$, $R_{\mathrm{long}}$, $H_s$, and $H_l$ are the surface net solar radiation, surface net longwave radiation, surface net sensible heat flux, and surface net latent heat flux, respectively. Frictional heating is ignored in $F_S$.

For global sum, the divergence term in equation (\ref{sec2_eq12}) is zero, and the global sum of the tendency of $A$ is balanced by its energy sources and sinks, subject to a residual term:

\begin{equation}\label{sec2_eq14}
\langle R_T \rangle - \langle F_S \rangle -\left\langle\frac{\partial A}{\partial t}\right\rangle = \epsilon_A
\end{equation}

\noindent
Here, the net energy flux is computed as $\langle R_T \rangle - \langle F_S \rangle$ because both terms have downward as positive. The downward on the top of the atmosphere means the energy goes ``into'' the atmosphere, but the downward on the surface mean the energy ``leaves'' the atmosphere. This is different from equation \ref{sec2_eq9} where both sources and sinks are at the surface.

The air temperature ($T$) can be corrected to ensure thermal energy ($C_pT$) closes the energy budget, forcing $\epsilon_A=0$:

\begin{equation}\label{sec2_eq15}
\begin{aligned}
\langle A^*\left(t_1\right) \rangle = \langle A\left(t_0\right)\rangle + {\Delta t}\langle R_T \rangle - \langle F_S \rangle, \quad \gamma = \frac{\langle A^*\left(t_1\right) \rangle}{\langle A\left(t_1\right)\rangle} \\
T^*\left(t_1\right) = \gamma T\left(t_1\right) + \frac{\gamma-1}{C_p}\left[L_v q\left(t_1\right)+\Phi_s+k\left(t_1\right)\right]
\end{aligned}
\end{equation}

\noindent
Where $\langle A^*\left(t_1\right) \rangle$ is the corrected global sum of total atmospheric energy, $\gamma$ is the multiplicative correction ratio. The same $\gamma$ is applied to $T$ at all grid cells and pressure levels.

\section*{Supplemental Material} % Custom title

% Reset figure counter and redefine numbering format
\setcounter{figure}{0}
\renewcommand{\thefigure}{\arabic{figure}S}

\subsection{Global Spectra}\label{S00}

The verification of the global energy spectrum is computed using spherical harmonic transforms. For a given forecasted or analyzed field $F\left(\phi, \lambda\right)$, it can be represented using spherical harmonic functions $Y\left(\phi, \lambda\right)$ as orthonormal basis and spherical harmonic coefficients ($a$):

\begin{equation}
F\left(\phi, \lambda\right) = \sum_{l=0}^{l_{\mathrm{max}}}{\sum_{m=-l}^l{a_l^m Y_l^m\left(\phi, \lambda\right)}}
\end{equation}

\noindent
Where degree $l$ represents the total angular frequency of $Y$. $m$ is the zonal wave number. The energy spectrum of $F$ at a given $m$ is the sum of magnitudes of $a$ in all degrees with $l\geq m$:

\begin{equation}
P\left(m\right) = \sum_{l\geq m}{\left\|a_l^{m}\right\|^2}
\end{equation}

The kinetic energy ($\mathrm{m^2\cdot s^{-2}}$) and potential temperature energy ($\mathrm{K^2}$) spectrum on 500 hPa pressure level were computed and as functions of $m$, $t_i$, and $t_l$. The result is averaged on $t_i$. Comparing $P\left(m\right)$ on forecasts and the ERA5 target, the ability of weather prediction models to represent the energy transfer across scales can be verified. In addition, the energy spectrum provides a measure of the effective resolution of AI NWP models, which helps identify the smoothing effect of neural-network-based computations and model training.

\begin{figure}[H]
    \centering
    \includegraphics[width=0.9\columnwidth]{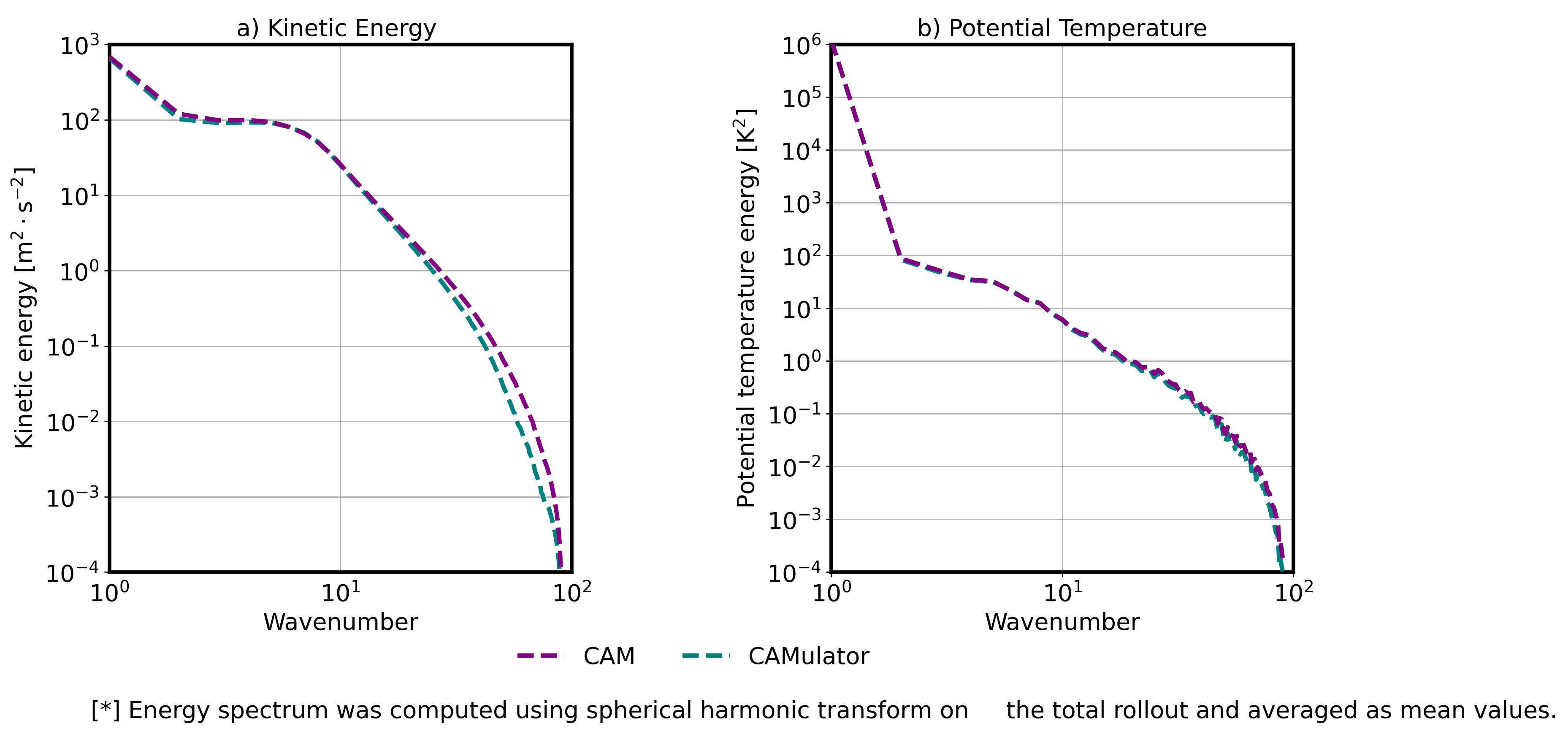}
    \caption{Global mean kinetic energy (a) and potential temperature (b) energy spectra for CAMulator (teal), and CAM6 (Purple) at 500 hPa calculated for years 1979-2010.}
    \label{fig:Spectra}
\end{figure}

\subsection{Annual Climatolgical RMSE}

\begin{figure}[H]
    \centering
    \includegraphics[width=0.9\columnwidth]{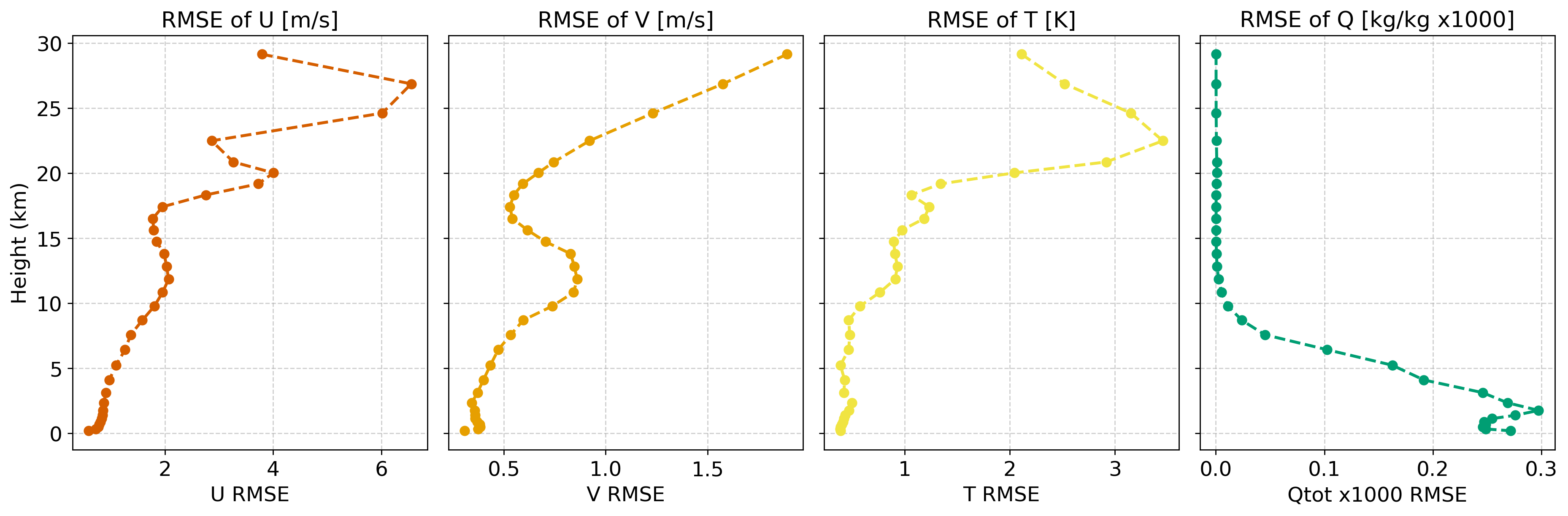}
    \caption{Annual climatological RMSE by level for U V T and Q for CAMulator simulation run from 1979-2010 computed against the CAM6 climatology}
    \label{fig:RMSE_by_Lev}
\end{figure}

\begin{figure}[H]
    \centering
    \includegraphics[width=0.9\columnwidth]{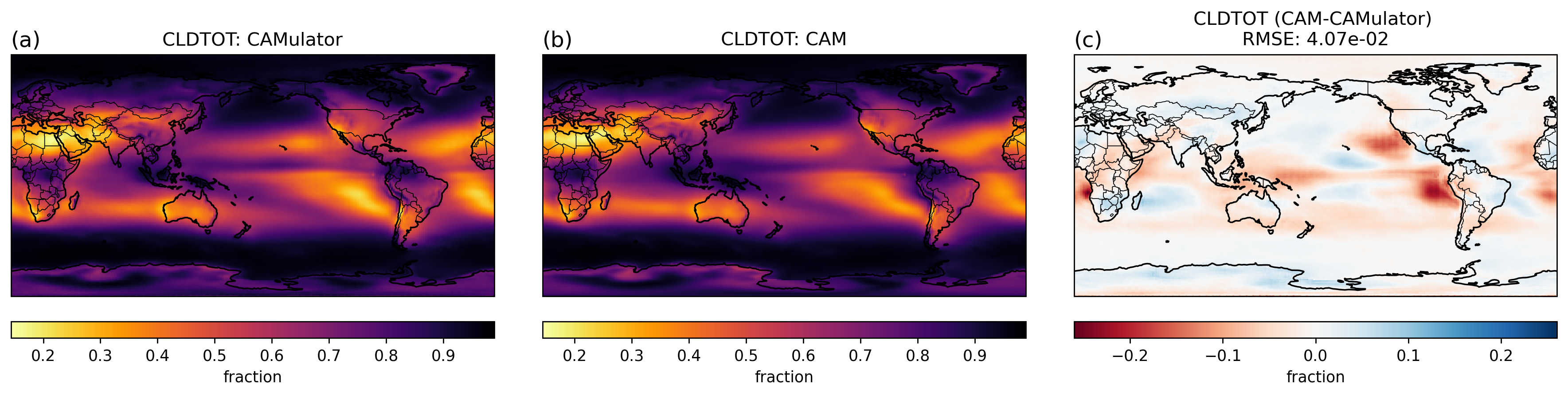}
    \caption{Annual climatology of the total cloud cover fraction in CAMulator (a), CAM6 (b) and the difference (c)}
    \label{fig:CLDTOT}
\end{figure}

\begin{figure}[H]
    \centering
    \includegraphics[width=0.9\columnwidth]{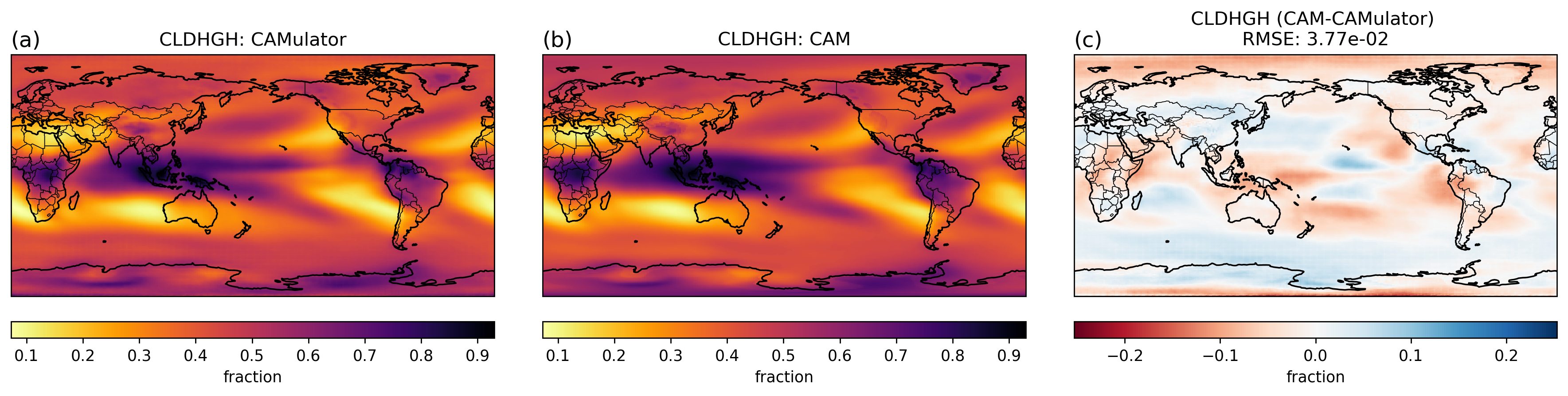}
    \caption{As in \ref{fig:CLDTOT} but for high cloud fraction}
    \label{fig:CLDHGH_RMSE}
\end{figure}

\begin{figure}[H]
    \centering
    \includegraphics[width=0.9\columnwidth]{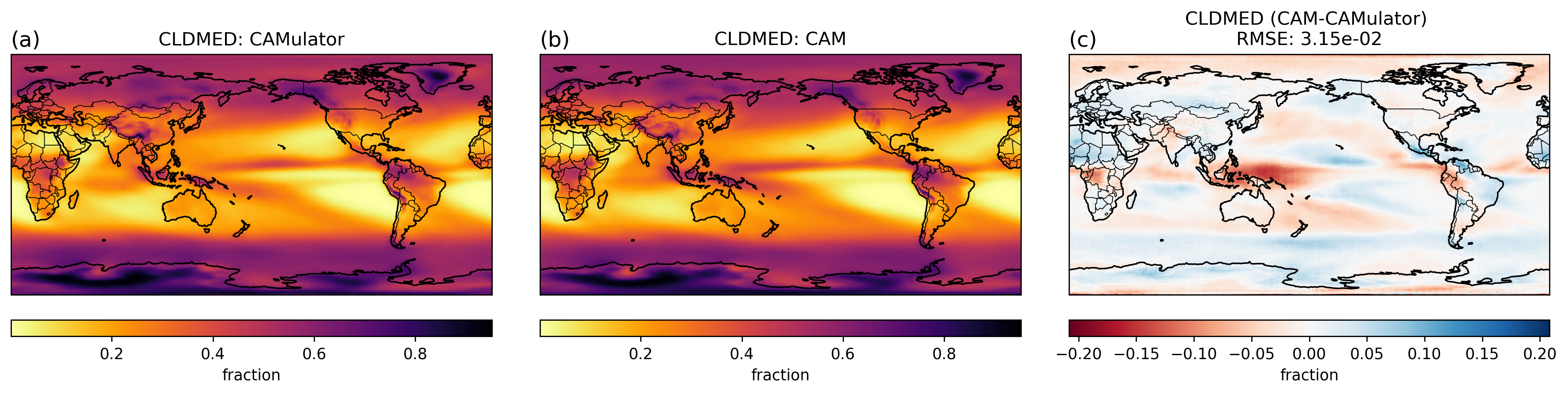}
    \caption{As in \ref{fig:CLDTOT} but for medium cloud fraction}
    \label{fig:CLDMED_RMSE}
\end{figure}

\begin{figure}[H]
    \centering
    \includegraphics[width=0.9\columnwidth]{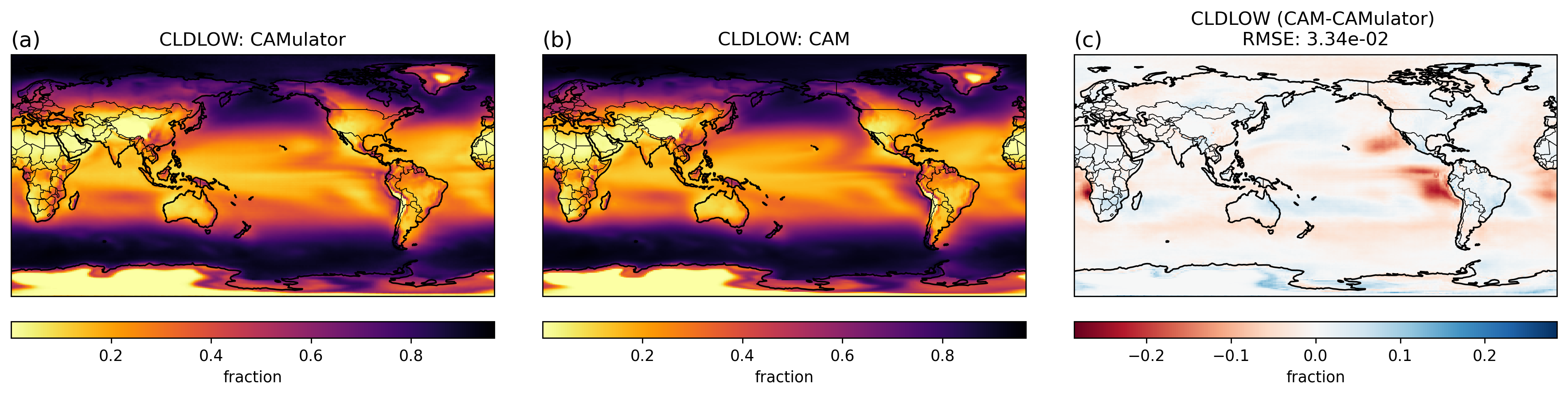}
    \caption{As in \ref{fig:CLDTOT} but for low cloud fraction}
    \label{fig:CLDLOW_RMSE}
\end{figure}

\begin{figure}[H]
    \centering
    \includegraphics[width=0.9\columnwidth]{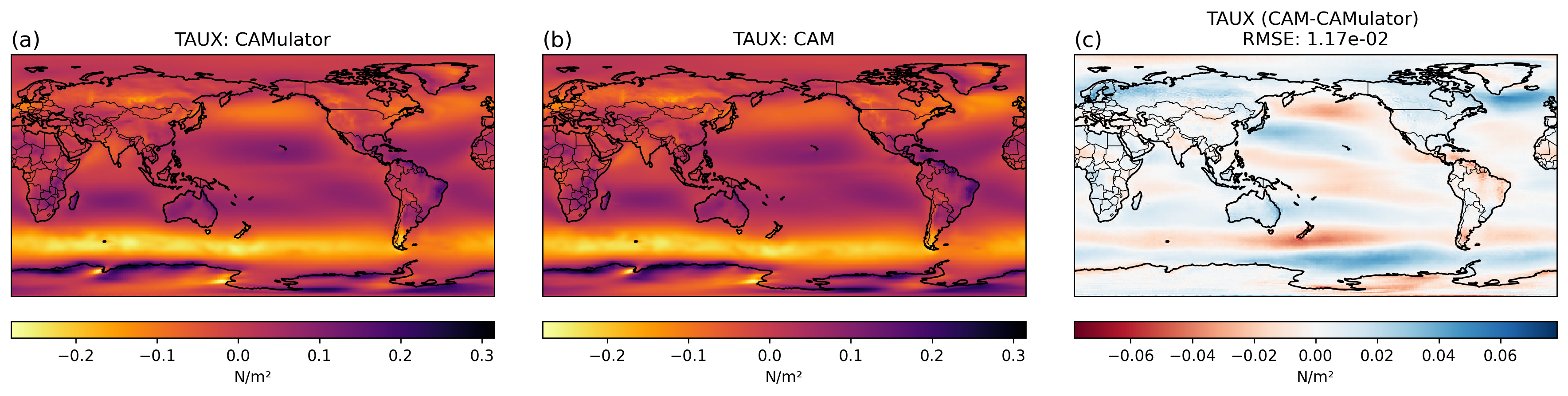}
    \caption{As in \ref{fig:CLDTOT} but for zonal surface wind stress}
    \label{fig:TAUX_RMSE}
\end{figure}

\begin{figure}[H]
    \centering
    \includegraphics[width=0.9\columnwidth]{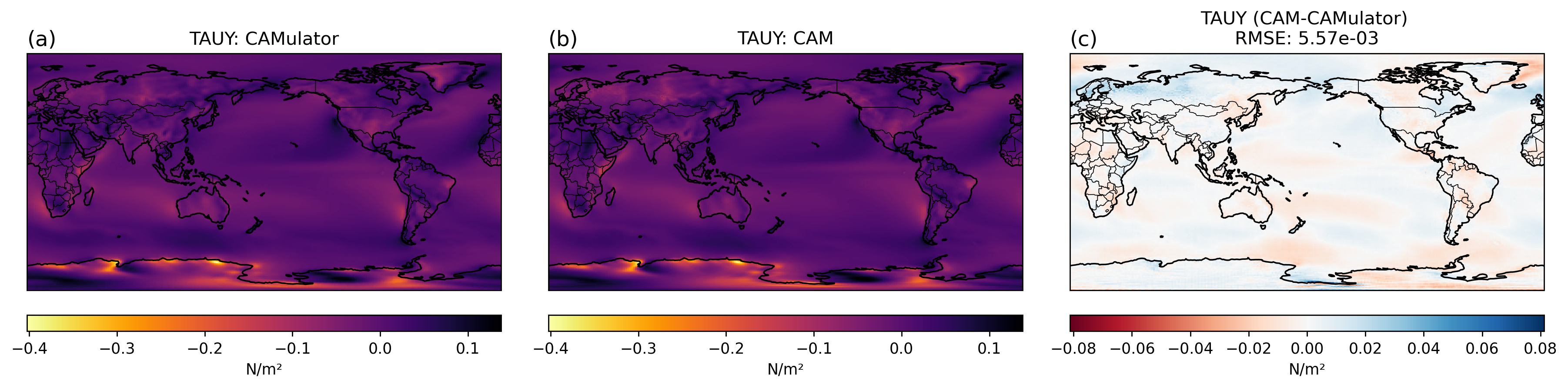}
    \caption{As in \ref{fig:CLDTOT} but for meridional surface wind stress}
    \label{fig:TAUY_RMSE}
\end{figure}

\begin{figure}[H]
    \centering
    \includegraphics[width=0.9\columnwidth]{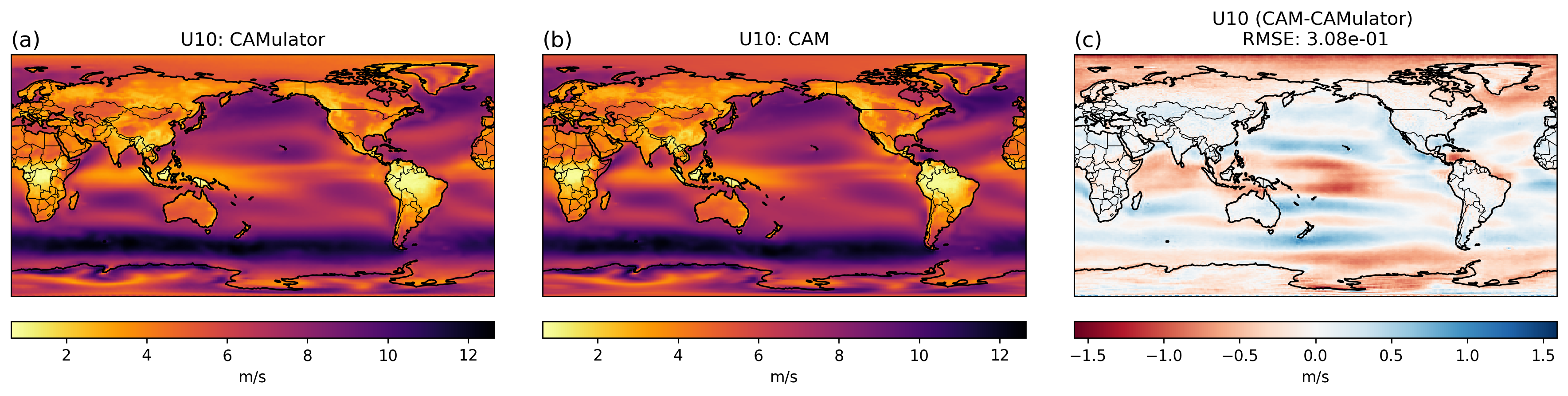}
    \caption{As in \ref{fig:CLDTOT} but for 10-meter wind magnitude}
    \label{fig:U10_RMSE}
\end{figure}

\begin{figure}[H]
    \centering
    \includegraphics[width=0.9\columnwidth]{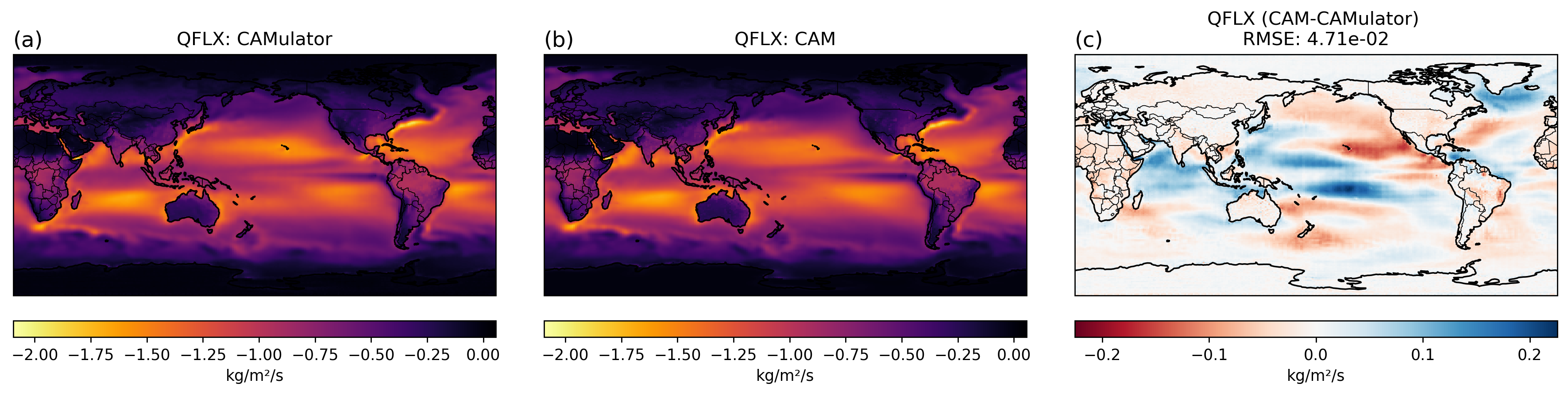}
    \caption{As in \ref{fig:CLDTOT} but for surface evaporation}
    \label{fig:QFLX_RMSE}
\end{figure}

\begin{figure}[H]
    \centering
    \includegraphics[width=0.9\columnwidth]{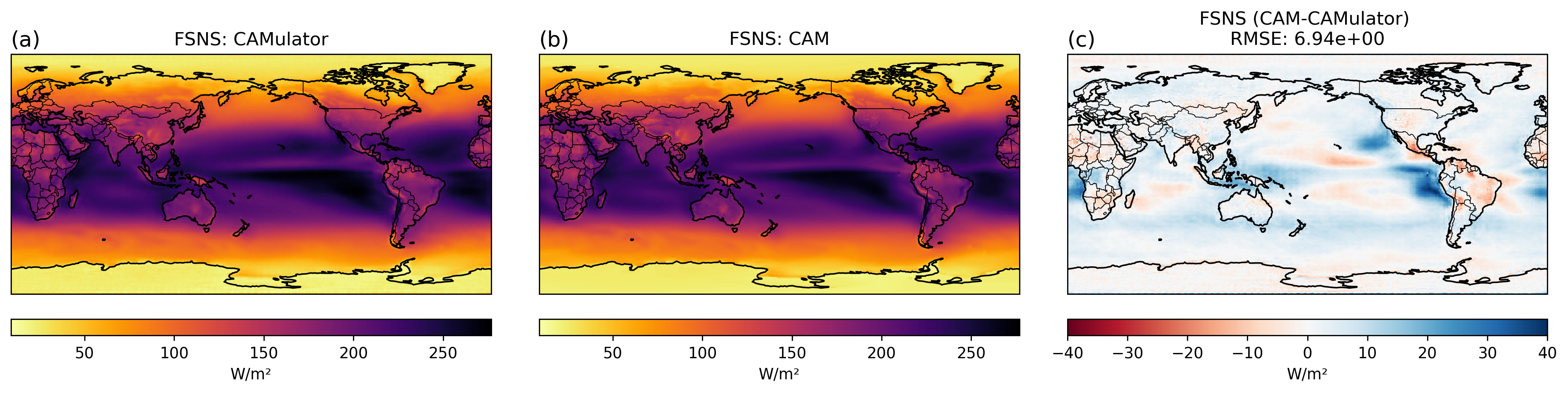}
    \caption{As in \ref{fig:CLDTOT} but for near-surface shortwave radiation}
    \label{fig:FSNS_RMSE}
\end{figure}

\begin{figure}[H]
    \centering
    \includegraphics[width=0.9\columnwidth]{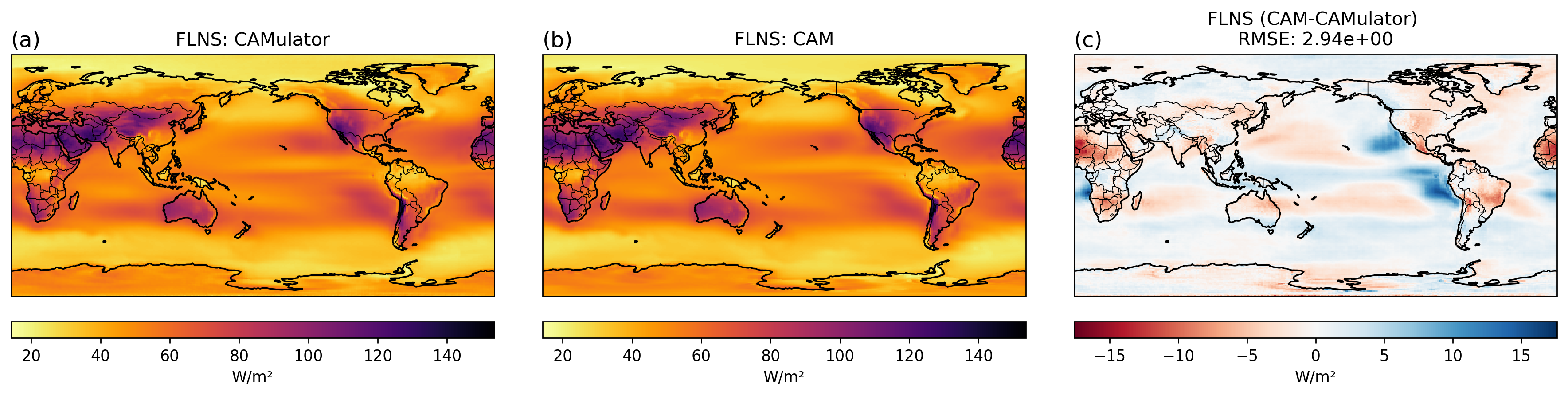}
    \caption{As in \ref{fig:CLDTOT} but for near-surface longwave radiation}
    \label{fig:FLNS_RMSE}
\end{figure}

\begin{figure}[H]
    \centering
    \includegraphics[width=0.9\columnwidth]{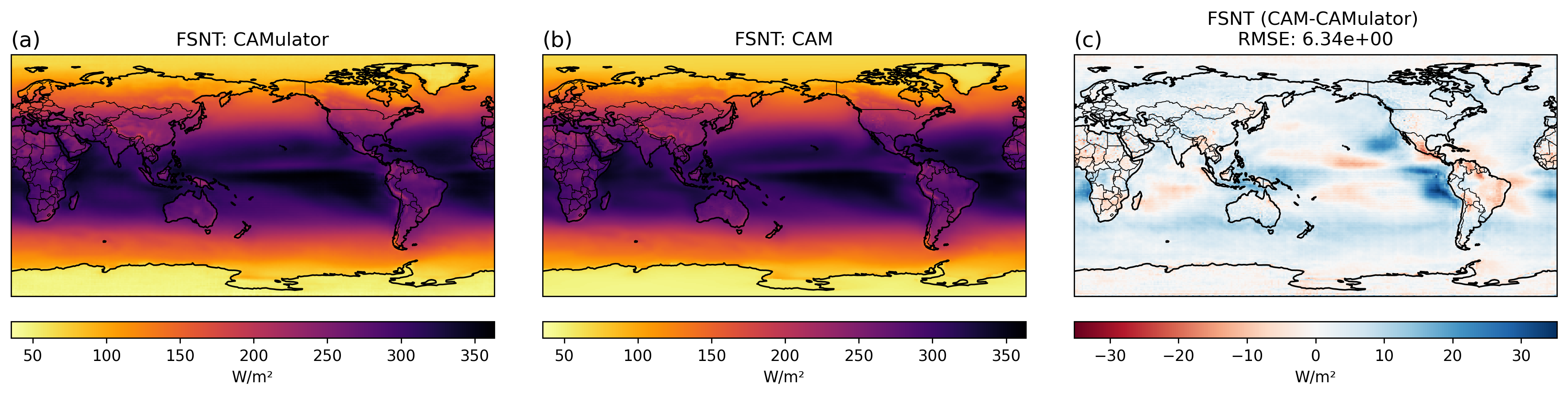}
    \caption{As in \ref{fig:CLDTOT} but for model top shortwave radiation}
    \label{fig:FSNT_RMSE}
\end{figure}

\begin{figure}[H]
    \centering
    \includegraphics[width=0.9\columnwidth]{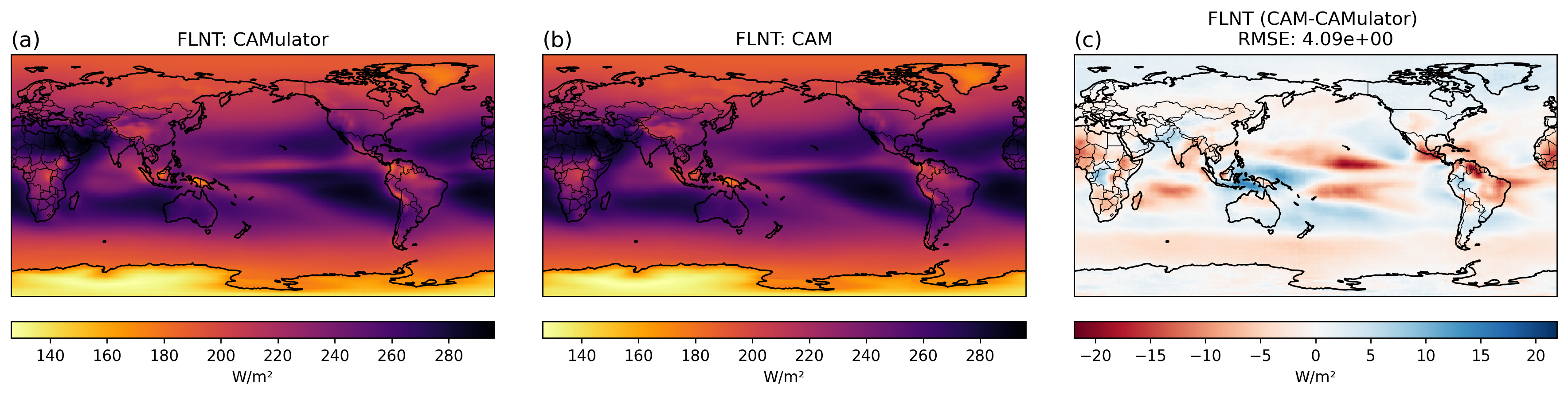}
    \caption{As in \ref{fig:CLDTOT} but for model top longwave radiation}
    \label{fig:FLNT_RMSE}
\end{figure}

\begin{figure}[H]
    \centering
    \includegraphics[width=0.9\columnwidth]{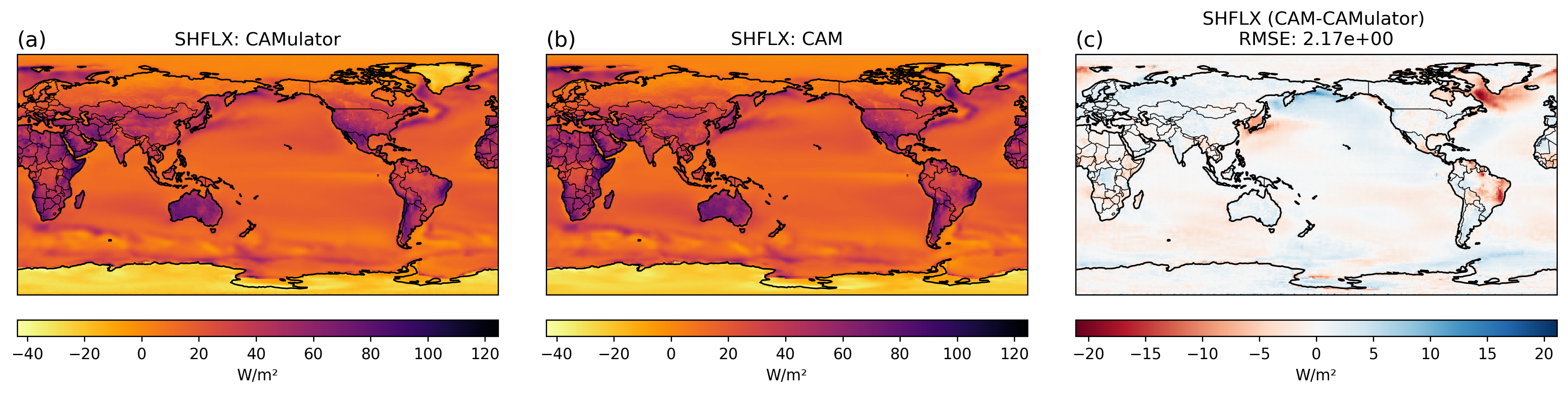}
    \caption{As in \ref{fig:CLDTOT} but for surface heat flux}
    \label{fig:SHFLX_RMSE}
\end{figure}

\begin{figure}[H]
    \centering
    \includegraphics[width=0.9\columnwidth]{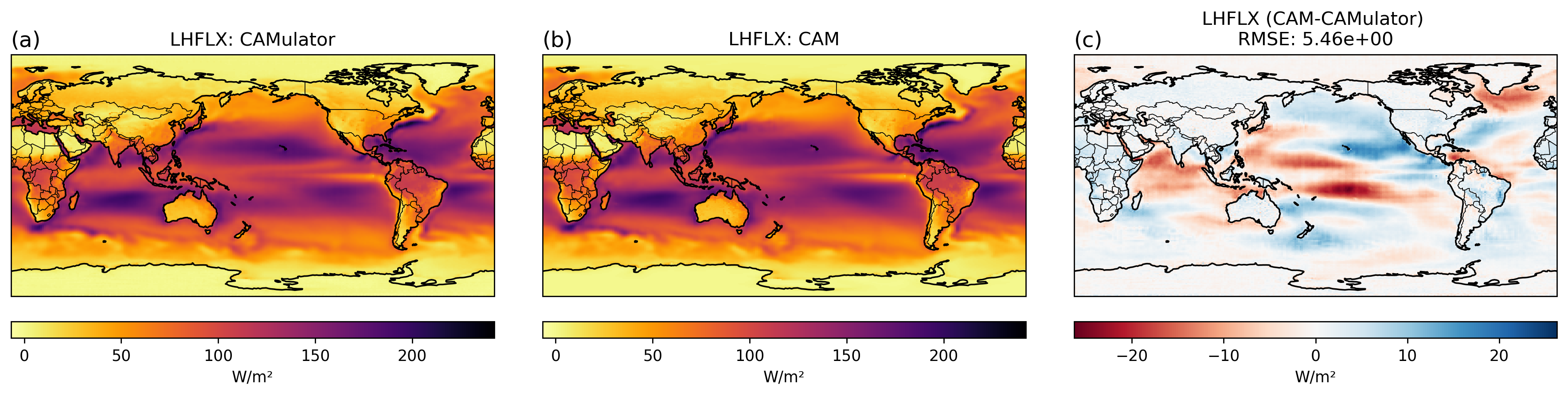}
    \caption{As in \ref{fig:CLDTOT} but for latent heat flux}
    \label{fig:LHFLX_RMSE}
\end{figure}

\begin{figure}[H]
    \centering
    \includegraphics[width=\columnwidth]{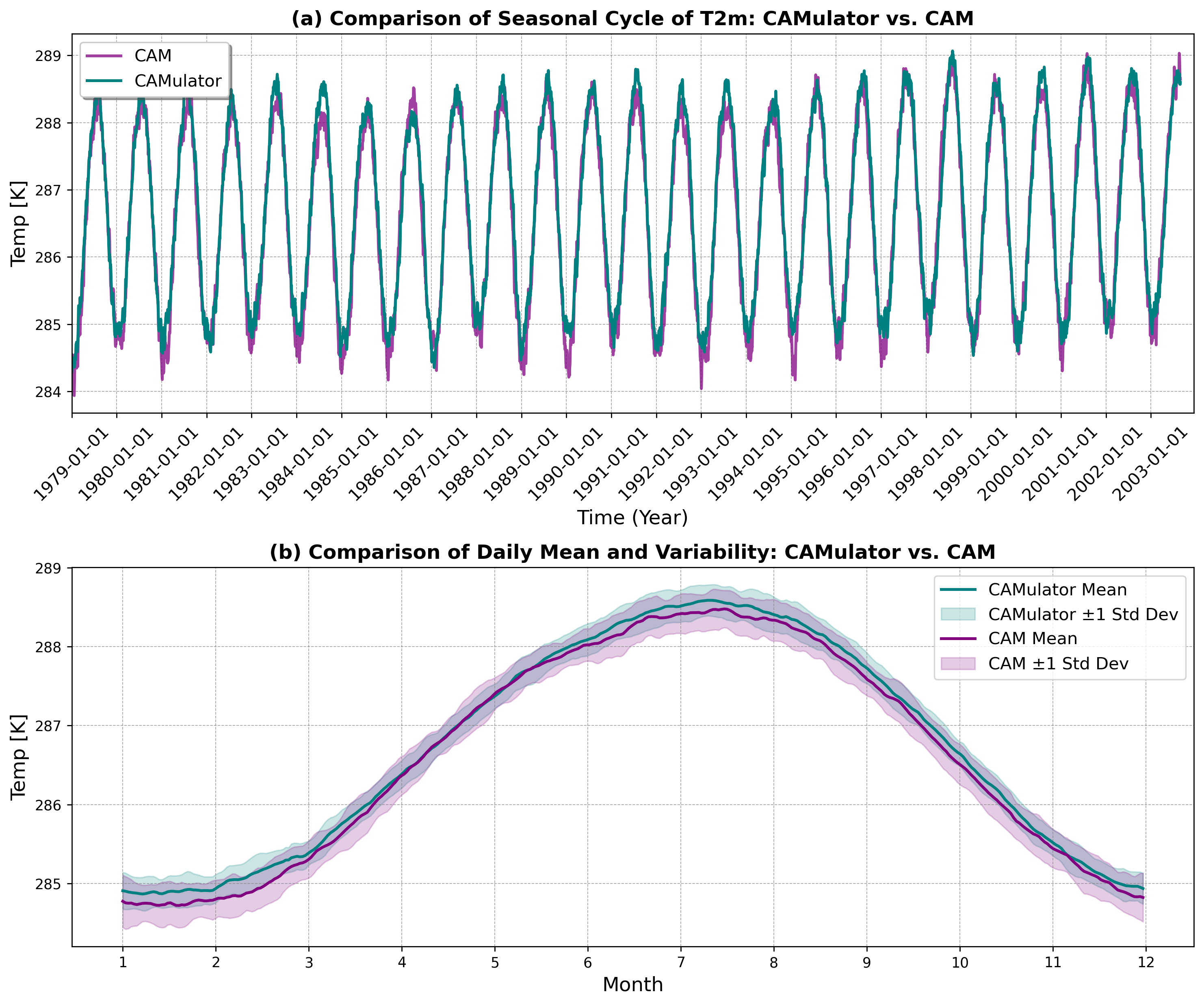}
    \caption{
    Time series of the latitude-weighted global average two-meter temperature (T2m) from CAM6 (purple) and CAMulator (teal). (Top) Seasonal cycle of T2m from 1979 to 2003, showing interannual variations. (Bottom) Climatological annual cycle of T2m, computed as the multi-year mean, with ±1 standard deviation shading representing interannual variability in CAM6 (purple) and CAMulator (teal). The latitude-weighting accounts for the cosine of latitude to ensure an accurate global mean representation.}
    \label{fig:Annual_T2m_Cycle}
\end{figure}

%Bibliography
\bibliographystyle{plainnat}  
\bibliography{references}

\end{document}